\documentclass[letterpaper,twocolumn,10pt]{article}
\usepackage{sty/usenix-2020-09}

\usepackage{tikz}
\usepackage{amsmath}
\usepackage{amssymb}
\usepackage{bytefield}
\usepackage{booktabs}
\usepackage{subcaption}
\usepackage{multirow}
\usepackage[table]{xcolor}
\usepackage{wrapfig}
\usepackage{pifont}

\usepackage{filecontents}

\usepackage{tikz}

\newcommand{\sysname}{ACORN}
\newcommand{\eg}{{\em e.g.,}}

\newcommand{\purple}[1]{\textcolor{purple}{#1}}

\newcommand{\fp}{\vspace*{0.02in}\noindent}
\newcommand{\boldpara}[1]{\fp {\bf #1}}

\newcommand{\circled}[1]{\tikz[baseline=(char.base)]{
  \node[shape=circle,draw,inner sep=1pt] (char) {#1};}}

\begin{document}

\date{}

\title{\Large \bf Towards Practical and Usable In-network Classification}

\author{
{\rm Di Zhu}\\
University of Virginia
\and
{\rm Jianxi Chen}\\
University of Virginia
\and
{\rm Hyojoon Kim}\\
University of Virginia
} 


\maketitle

\begin{abstract}
In-network machine learning enables real-time classification directly on network hardware, offering consistently low inference latency. However, current solutions are limited by strict hardware constraints, scarce on-device resources, and poor usability---making them impractical for ML developers and cloud operators. To this end, we propose \sysname{}, an end-to-end system that automates the distributed deployment of practical machine learning models across the network. \sysname{} provides a fully automated pipeline that loads and deploys Python ML models on network devices using an optimized deployment plan from an ILP planner. To support larger models under hardware constraints and allow runtime programmability, \sysname{} adopts a novel data plane representation for Decision Tree, Random Forest, and Support Vector Machine models. We implement \sysname{} prototype in P4 and run it on real programmable hardware. Our evaluation shows \sysname{} can deploy classification ML models with 2--4x more features than state-of-the-art solutions, while imposing negligible overhead on network performance and traffic.
We will make our data plane program, model translator, optimizer, and all related scripts publicly available. 
\end{abstract}
\section{Introduction}

With the emergence of flexible networking hardware~\cite{bosshart2013forwarding} and data plane level programming languages~\cite{bosshart2014p4}, networks now support more functionalities without sacrificing packet processing rates.
This trend, along with the rise of commodity programmable network devices, creates the opportunity for in-network computing, the idea of offloading computing tasks from end hosts to network devices. In-network computing offers significant performance benefits, such as line-rate throughput, lower latency, power efficiency, and load reduction, and we have already witnessed its success in many applications (\eg{} \cite{liu2017incbricks,ghasemi2017dapper,jin2017netcache,ben2020pint,liu2021jaqen,jin2018netchain,zhang2020poseidon}).

In particular, the growing demand for Artificial Intelligence (AI) and Machine Learning (ML) workloads is driving the need for innovative and non-traditional computing devices.
However, delegating machine learning inference jobs to the network faces many practical challenges. 
(i) Different AI/ML models have different features and structures, which generally require a customized, model-specific network-level program. This makes it challenging to make in-network AI runtime programmable. 
\textit{How can we make in-network AI support a broad range of models while supporting runtime programmability to minimize downtime whenever a model update happens?}
(ii) Practically usable AI/ML  models do not fit into a single programmable data plane. Yet, current P4 program compilers are optimized for deployments on a single data plane, not across two or more. For larger models, the slicing and distribution of a network-level program is done manually where even experts rely on trial-and-error. This process shifts the focus from optimizing performance to simply achieving functional deployment. \textit{How do we automate the process while ensuring the lowest inference latency and minimizing total resource usage on network hardware?}
(iii) The process of translating, compiling, and deploying AI/ML models on programmable data planes is currently tedious, error-prone, and labor-intensive. AI/ML applications are typically written in high-level programming languages such as Python while applications for programmable switches are developed in lower-level languages such as P4~\cite{bosshart2014p4}. The transformation process requires extensive knowledge of programmable packet processing and proficiency in low-level languages.
\textit{How do we democratize in-network AI/ML for practitioners, like AI/ML developers or cloud operators, without needing to gain expertise in pipeline programming?}

In this paper, we propose \sysname{}, an automated and distributed computing system for in-network machine learning classification applications. To make our system friendly towards not only network operators but also application developers, we designed an automated pipeline that translates Python ML models, optimizes for a deployment plan, and deploys the model on the data plane. We proposed a new data plane design for decision trees and random forests that enables the deployment of bigger models. Our data plane is carefully designed to support multiple models and is runtime programmable at the same time. In summary, our contributions are:

\boldpara{Fully automatic control plane pipeline.}
We present a fully automatic control plane pipeline that converts Python ML programs into table representations, generates an optimal deployment plan, and deploys the program on the network.

\boldpara{Novel representation for tree-based models.}
We propose a novel representation for tree-based models that supports up to 60 features and 32 depth of trees that satisfies the needs for both traffic analysis and general ML classification tasks.

\boldpara{Runtime configurable data plane.} 
We present a runtime reconfigurable data plane for distributed in-network ML systems that enables switching between ML applications without recompiling networking devices, while facilitating the communication between devices in the network. 
This is achieved by predefining five reusable match-action tables types that perform core operations common to many ML workloads.

\section{Background and Motivation}
We introduce background information on in-network ML and the limitations of current state-of-the-art systems.
\subsection{In-network computing and ML}

\boldpara{In-network ML:}
Previous works, like pForest~\cite{busse2019pforest} and IIsy~\cite{xiong2019switches}, have pioneered the research on in-network ML. They proposed frameworks for in-network classification with traditional machine learning models by mapping them to a match-action pipeline. Planter~\cite{zheng2024planter} improves upon IIsy and proposes to automate in-network classification by incorporating ML models into modules.
The main advantage of running ML-powered classification workloads on programmable data planes, such as SmartNICs and switches, is its \textit{significantly lower inference latency}. As data planes are designed for extremely fast packet processing, matching, and actions, match-action table-based classification can complete the task several orders of magnitude faster than running the same job on a general-purpose CPU machine~\cite{swamy2022taurus, zhang2020poseidon}.

\boldpara{In-network computing and resource constraints:}
Data plane programs are usually compiled into several match-action tables (MATs) in a programmable data plane. Such a table matches packet headers and performs actions based on a matching result. However, programmable network devices have constrained resources because they are designed to be performance-driven and operate at line rate. For example, the  Intel Tofino 2 switches~\cite{Tofino} offer a bandwidth up to O(10) Tb/s-scale, but only have O(10) MB-scale memory~\cite{kim2020tea} and less than two dozens pipeline stages. The scarce resource capacity forbids the deployment of large applications or concurrent deployment of multiple programs that consume large amounts of resources. This is especially the case for ML-related applications, as the number of parameters an ML model can have is bounded by the available resources on the machine, the more parameters usually lead to models with better performance as they can model more complex relations within the dataset. Thus, there has been a rising interest in distributed in-network computing, to achieve better resource utilization and higher performance.

\boldpara{Distributed in-network ML:}
More recent works adopt \textit{distributed computing}, enabling distributed in-network ML. 
There are two distributed in-network computing systems that support ML programs; DINC~\cite{zheng2023dinc} and DUNE~\cite{butun2025dune}. Although they have achieved success and showed promise, there is still a gap between these state-of-the-art systems and practical distributed in-network systems for ML applications.

\subsection{The Gaps in Distributed In-network ML}

\smallskip
\noindent
\textbf{Poor usability for application developers.} Both DINC and DUNE require P4 programs as inputs, but hardware-level languages such as P4 are difficult for application developers to learn and use. There is also a limited level of automation in both work. In DINC~\cite{zheng2023dinc}, the program needs to be partitioned, optimized, and deployed \textit{manually}. DUNE~\cite{butun2025dune} improves DINC by automatically disaggregating ML models, but this procedure must be repeated whenever there is a change in the model with manual configurations, which requires expertise. 

\boldpara{No support for runtime programmability.} Offloading different ML applications to the network involves deploying different programs on the networking devices, and this process could seriously affect normal packet forwarding. Although some architecture, such as Intel Tofino, could reduce the compilation time to as little as O(50ms)~\cite{FastRefresh}, considering all the devices that need recompilations, the total impact could not be underestimated. Thus, we argue that runtime programmability is an important feature for distributed INC systems.

\smallskip
\noindent
\textbf{Limited supported number of features.} Many datasets used in building domain-specific systems, such as IDS systems, contain 20 to more than 100 features~\cite{cicids2017, unswnb15, dong2023horuseye, nsl-kdd}. However, current distributed in-network systems for ML were evaluated using small datasets, or a small subset of features from bigger datasets. For example, for Random Forest models, DINC uses a model with 6 stages, suggesting that the dataset used is a 6-feature dataset; DUNE extracts flow-level features from two datasets, UNSW-IoT~\cite{unsw-iot} and TON-IoT~\cite{ton_iot}, but it is unclear how any features were used in the evaluations. According to our experiments (discussed in \S\ref{sec:eval}), both systems encountered issues when the number of features increase.
\section{\sysname{}: Design Principles and Workflow}

\subsection{Design Principles and Key Techniques}
\sysname{} is system that automatically translates a system-level ML program
into a network-level program, generates an optimal deployment plan, and distributes
the target workload across programmable data planes.

\boldpara{Novel representation for tree-based models.}
We propose a novel design and representation for tree-based models (\eg{} decision trees, random forests) that supports up to 60 features and 32 depth of trees. Together with our distributed deployment capability, \sysname{} can support bigger and more practical ML models than previous solutions, ranging from traffic analysis to general ML classification tasks. In this paper, we demonstrate \sysname{}'s usability and applicability with 12 different models and workloads (\S~\ref{sec:trans}).

\boldpara{Fully automatic optimal deployment.}
A user submits a single input to \sysname{}: a trained ML model. \sysname{} automatically translates the model into a series of table representations. Then \sysname{} runs its optimizer to find the best deployment plan and modifies the table entries on the data planes. 
Our optimizer accounts for factors often overlooked in prior work, including per-packet processing time, hop count, and network overhead such as packet size inflation.
Query packets are prepared and sent directly from the host of the user who submitted the model (\S~\ref{sec:trans}, \S~\ref{sec:opt}).

\boldpara{Runtime programmable.}
We design a runtime reconfigurable \textit{template data plane program} that is compiled and deployed \textit{once} in the target network. A user can switch and deploy a new ML by flushing and installing new flow table entries in predefined match-action tables in the P4 program while leaving the forwarding table entries intact. \sysname{} does not require recompiling and reloading networking devices, facilitating continuous communication in the network. To achieve this, we create five table types that are capable of covering a broad range of ML use cases: \texttt{DT\_layer}, \texttt{DT\_predict}, \texttt{Multitree\_voting}, \texttt{SVM\_mul}, and \texttt{SVM\_predict} (\S~\ref{sec:run_prog}).

\subsection{Workflow}

\begin{figure}[t]
\centering
\includegraphics[width=\linewidth]{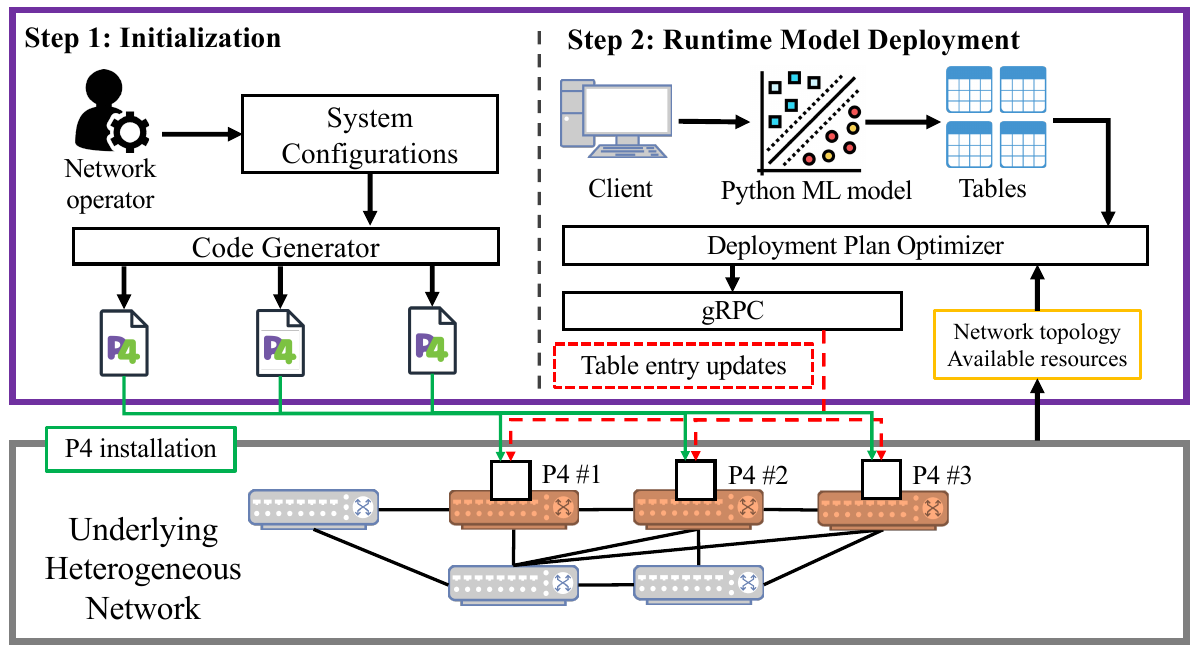}
\caption{An overview of \sysname{}'s workflow.}
\label{fig:overview}
\end{figure}

Figure~\ref{fig:overview} shows the high-level architecture and workflow of the proposed system. The workflow consists of two phases. The first step (initialization) creates and deploys a unified P4 program template with modular match-action tables.

The network can be heterogeneous---the P4 programs are only deployed and loaded on P4-programmable data planes in the given network, and \sysname{} control plane, including the deployment optimizer, is aware of this. 

At this point, the underlying network is ready to receive ML classification tasks, or the second step. Currently, \sysname{} supports ML models built using Python with \texttt{scikit-learn} library. Model parameters and other important information about the model are sent to the system using an API. Based on the information, the code translator translates the model into a series of match-action tables. The deployment plan optimizer uses the number of table entries, network topology, and the available resources within the network to generate the optimal strategy for deploying these tables. However, instead of (re)installing new tables and recompiling the switches, our proposed system achieves runtime programmability by updating the entries in predefined tables. Finally, inference requests can be sent by clients, using specially-formatted \sysname{} packets. Appendix~\ref{appen:header} describe the format of a \sysname{} packet.

\section{Model Representation and Translation}
\label{sec:trans}
We currently support three types of models: decision trees, random forests (bagging-based forests), and support vector machines (SVMs).

\subsection{Decision Trees}
In decision trees, non-leaf nodes represent tests on the input data, branches represent outcomes of the tests, and the leaves represent class labels or predictions. Tree-based models are inherently easy to represent within programmable switches, as match-action tables (MATs) function on similar logic. For example, a non-leaf node can be thought of as a MAT, where the input data is the key, and the actions are the outcomes (branches). However, each decision tree may contain a large number of nodes, and mapping each node to a MAT will be unrealistic. Thus, it is crucial to design an efficient representation to enable the offloading of larger trees to the network. 

\smallskip
\noindent
\textbf{Previous representations.} Multiple previous works, including SwitchTree~\cite{lee2020switchtree}, pForest~\cite{busse2019pforest}, IIsy~\cite{zheng2024iisy}, and LEO~\cite{jafri2024leo}, have proposed different representations for decision trees. These representations are designed for single-device systems, and we could not simply adapt them to distributed systems due to their limitations. 
For instance, IIsy~\cite{zheng2024iisy} seeks to decouple tree depths from pipeline stages by flattening the decision paths. However, its resource consumption can become unrealistic~\cite{jafri2024leo}, which prohibits the deployment of models in some cases. For example, it requires more than $3*10^{11}$ TCAM entries to represent a decision tree trained on a reduced 10-feature Digits~\cite{sklearn_digits} dataset. LEO~\cite{jafri2024leo} solved this problem by dividing a decision tree into multiple sub-trees and performing multiplexing to save resources and improve pipeline stages usage. However, due to sub-tree multiplexing, multiple tables share the same stage, and the number of features that LEO supports is strictly limited to no more than 10 features with a sub-tree size of 3. 

\begin{figure}[t]
  \centering
  \includegraphics[width=\linewidth]{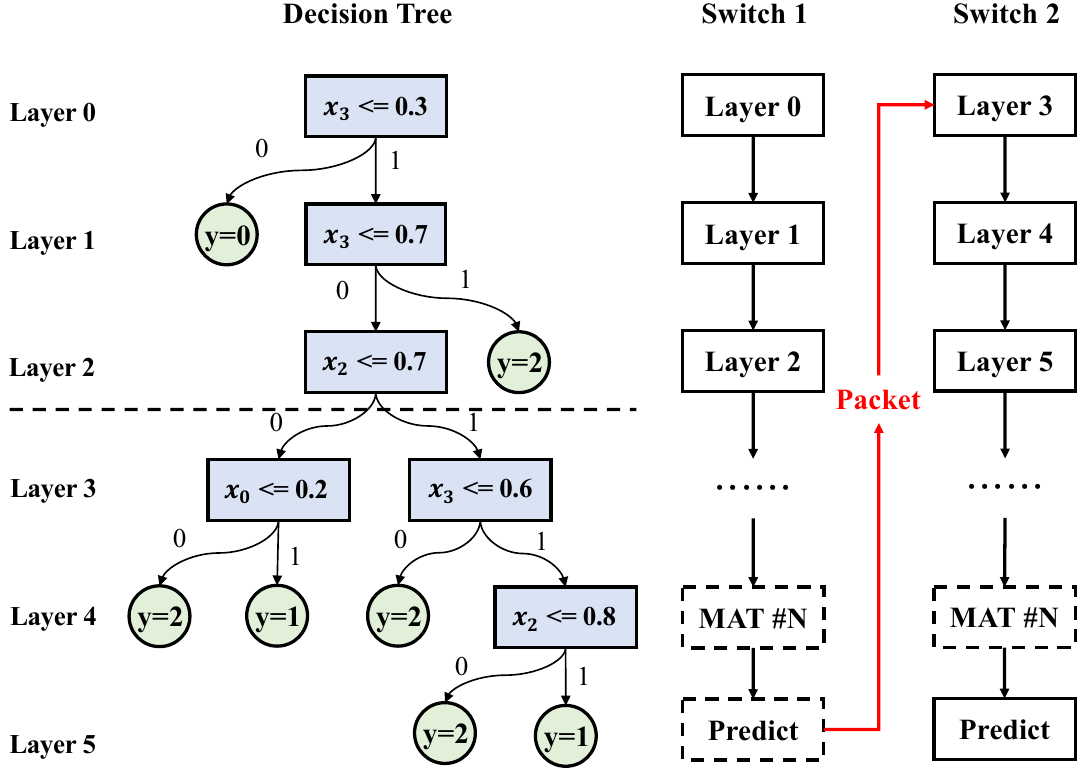}
  \caption{Distributing a decision tree (Iris dataset~\cite{iris_53}).}
  \label{fig:dt_iris_example}
\end{figure}

\boldpara{\sysname{}'s layer-based representation.} 
To this end, we adopt a layer-based strategy to represent trees by mapping one layer of trees to a pipeline stage. 
As shown in Figure \ref{fig:dt_iris_example}, layers 0, 1, and 2 are distributed to switch 1, while layers 3, 4, and 5 are installed at switch 2. At each layer, there is one MAT that matches on the pair of \texttt{(Feature, PrevStatusCode)} using ternary matching keys, where the \texttt{Feature} is the target feature to be tested. The \texttt{PrevStatusCode} is a 16 to 32 bits bitstring, in which each bit represent a decision from a previous layer. If a match is hit, the MAT will update the status code, embedded in the packet, to describe the action at each level, which is represented by a single bit---0 and 1 represent left and right, respectively. 
If a leaf node in the tree is reached before the packet traverses the entire pipeline, the packet passes through the remaining tables without triggering any actions, and the prediction table produces a label based on the status code.

\begin{figure}[t]
  \centering
  \includegraphics[width=\linewidth]{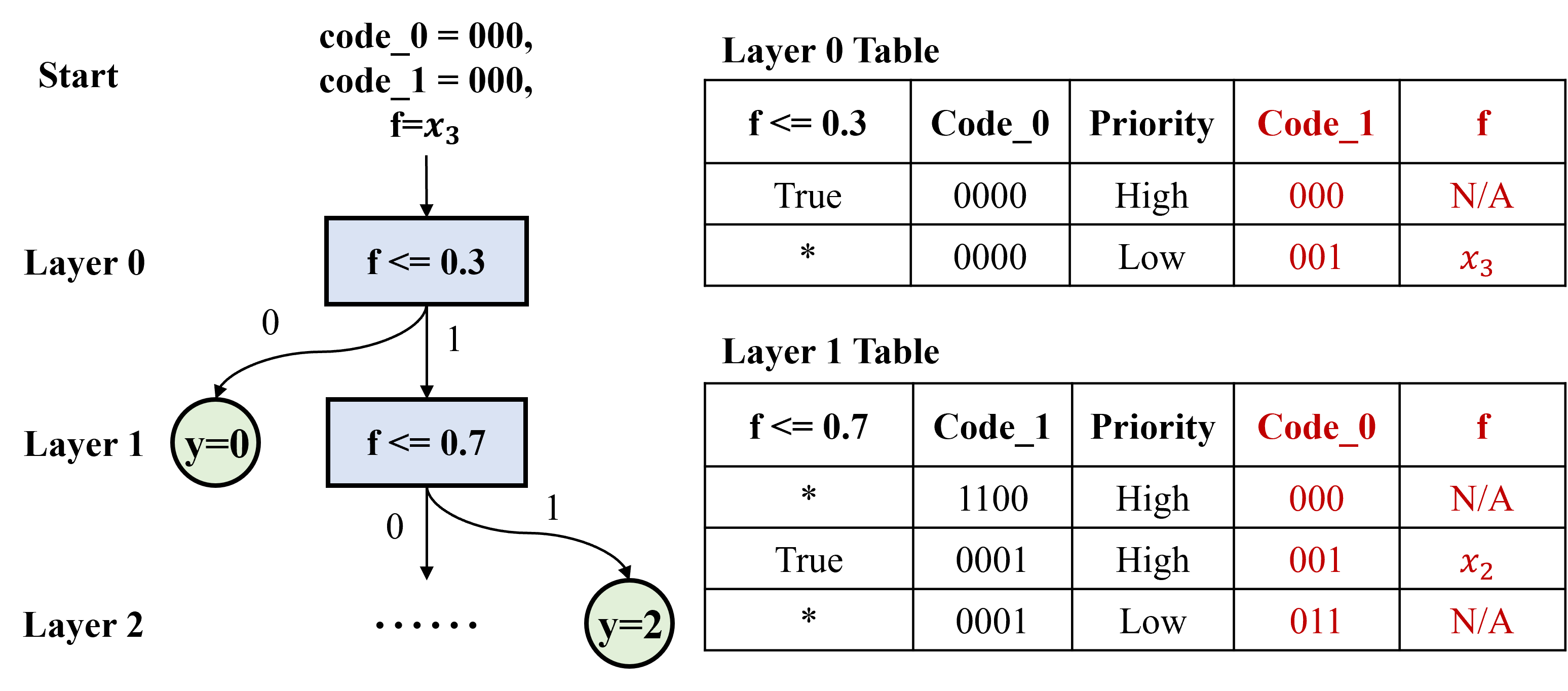}
  \caption{Encoding a decision tree in \sysname{}. On the tables, packet matches on the first three columns (black). Red columns are the actions.}
  \label{fig:dt_encoding}
\end{figure}

\boldpara{Example. }
Figure \ref{fig:dt_encoding} shows a detailed example. The switch first parses the packet and extracts three variables: \texttt{code\_0}, \texttt{code\_1}, and \texttt{f}. Variables \texttt{code\_0} and \texttt{code\_1} are status codes, where \texttt{code\_0} is matched in even-numbered layers, and \texttt{code\_1} is used as one of the keys in odd-numbered layers. Variable \texttt{f} indicates the feature that is going to be tested. At layer 0, if $x_3$ satisfies the condition, the table updates \texttt{code\_1} to 0, indicating that the data point will reach the leaf node at layer 1; if not, \texttt{code\_1} will be updated to 1, meaning that more MATs need to be triggered to make the prediction. The entry priority is used to reduce the number of table entries, thereby minimizing on-device resource usage. At layer 1, the table first checks if the data has reached a leaf node before checking if the target feature satisfies the condition. The computation continues until it reaches the end of the switch pipeline.

Our representation has several prominent advantages over previous representations. First, our representation can support a larger number of features without using an excessive amount of SRAM/TCAM resources. Second, although our layered representation of tree models uses more pipeline stages, it can support trees with depths of 32 or more.  For distributed systems like \sysname{}, we can compensate for this by using more devices in the network.

\subsection{Random Forest}
For multi-tree models, we focus on supporting random forest models similar to previous literature~\cite{gao2019adaptive, sharma2023estimating, bronzino2019inferring, xie2022mousika}.
A random forest contains multiple independent decision trees. During the inference process, each decision tree makes individual decisions, and an extra voting step combines all decisions and makes the final prediction.
\sysname{} translates random forests based on this observation, by decomposing the inference procedure into multiple decision trees, which can be translated based on the methods discussed above, and defines a final prediction table that combines the outputs of all decision trees and outputs the final prediction. Different combining schemes, such as majority voting and weighted summation, can all be represented as voting. In this way, we only need to define one table to satisfy the needs of random forest models without increasing resource usage and program complexity.

\subsection{Support Vector Machine}
Support Vector Machines (SVM) divide the feature space using several hyperplanes and classify data by finding the optimal hyperplane that maximizes the distance between two classes in the feature space.

\boldpara{Previous representations.} Previous work, such as IIsy~\cite{zheng2024iisy}, has proposed two ways of representing SVMs. The first way is to use one table per hyperplane, by matching the concatenation of all input features and returning the code of the hyperplane as the action. This method effectively reduces the complexity of the data plane program, but is strictly bounded by the width of the concatenation of all features. Yet, the number of concatenations could grow \textit{exponentially} as the number of features increases, eventually becoming too many for MATs to hold. For example, for an SVM with 4 features and 4 digit precision, the number of entries will be $(10 ^ 4)^4$, which is infeasible for most programmable data planes.

The second approach proposes using one table per feature, and each feature table outputs all intermediate multiplication results for this feature. For example, if an SVM model has 3 hyperplanes, the table for feature $x_i$ will output $x_i * w_{0i}$, $x_i * w_{1i}$, and $x_i * w_{2i}$. These intermediates will be added using the native addition operator for the final prediction. Although this method greatly reduces the number of table entries, all intermediate results have to be carried in the packet to other devices, which creates extra overhead. 

\begin{figure}[t]
  \centering
  \includegraphics[width=\linewidth]{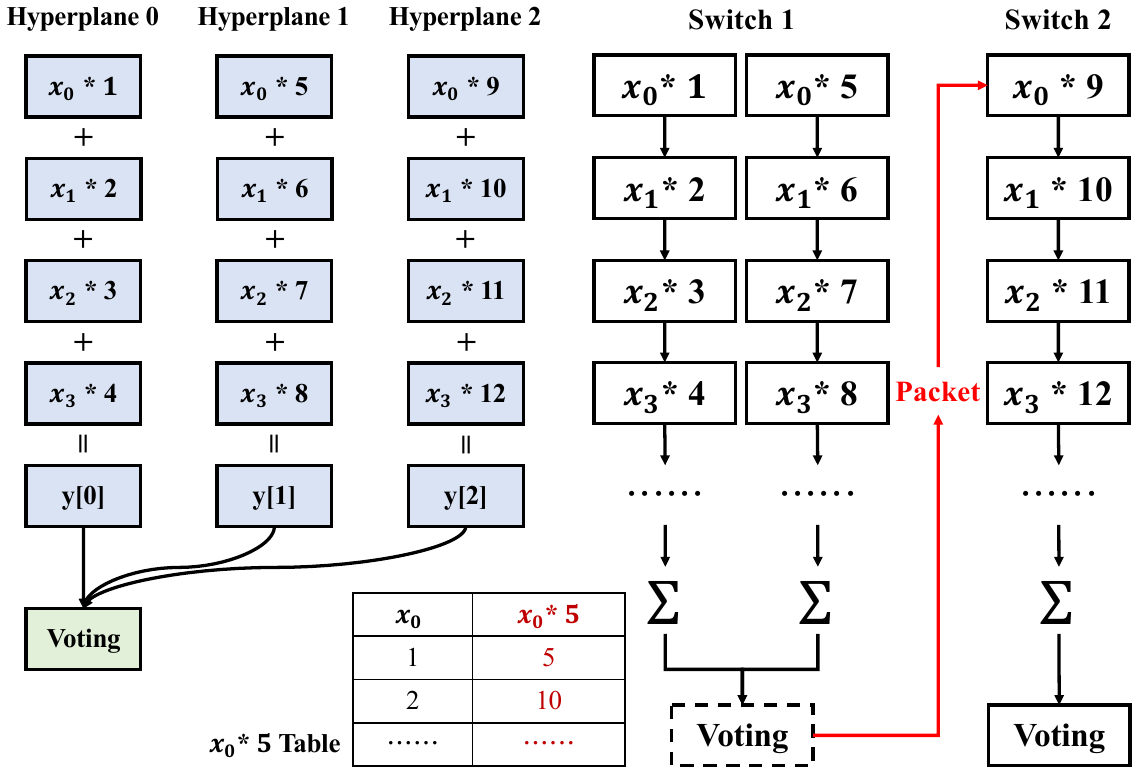}
  \caption{Encoding a support vector machine model.}
  \label{fig:svm_encoding}
\end{figure}

\boldpara{\sysname{} representation.}
\sysname{} uses a simplified design to represent SVMs. Since there could be multiple SVM pipelines in a switch, instead of using one table per feature, \sysname{} uses one table per multiplication to fully utilize the on-device resources. The hyperplane results are calculated using the native addition operator at the end of the pipeline. Since the final prediction only uses the sign of the hyperplane results, \sysname{} uses signed integers to store features and extracts the highest bits as the code for the hyperplanes. 
In each SVM pipeline, multiple multiplication tables can be placed in the same pipeline stage, improving parallelism and saving pipeline stages. 
\textbf{Example.} In Figure \ref{fig:svm_encoding}, the example SVM model is distributed across two switches; hyperplane 0 and 1 at switch 1 and hyperplane 2 at switch 2. The intermediate result from switch 1 is carried to switch 2 though packets, and switch 2 makes the final classification decision by votes.

\section{Optimization}
\label{sec:opt}
\sysname{} formulates a multi-objective optimization problem. 
In this section, we present the design of \sysname{}'s optimizer. 

\subsection{Assumption and Problem Definition}
We assume that the underlying network comprises end-host machines and network devices. Our optimizer and \sysname{} support a heterogeneous network with a mix of programmable and non-programmable devices.
Models can be deployed on programmable devices, whereas non-programmable devices only forward packets. 

The total set of supported ML models is $M=\{m_i\;|\;i\in[N_M]\}$. The following definition applies to each type of supported model $m$.
We define the network as an undirected graph $Network=(D, E)$, where $D=\{d_k \; | \; k \in [N_D]\}$ is the set of all devices on the network, and $E=\{e_{ij} \; | \; i,j \in [1, N_D] \; \cap \; \mathbb{Z}, i \neq j\}$ is the set of all links connecting these devices. Each device $d_k$ has three properties: (1) Programmability, indicated by $d_{k}^{p} \in \{0, 1\}$. A value of 1 indicates that device $d_k$ is loaded with an \sysname{} data plane; (2) The number of available stages $D_s$; (3) The size of available space for table entries in stage $i$, table $j$ is $d_{k}^{s}(i,j)$.

The target ML model is converted into a sequence of tables. Assume it requires $T_s$ stages with $T_m$ tables per stage. Each table $t_{i,j}, \text{where} \; i\in[T_s], j\in[T_m]$ requires $t_{i,j}^{e}$ entries.
Inference request packets start at the source device $d_{src}$, travel through the network while being processed, and arrive at the destination device $d_{dst}$. Devices $d_{src}$ and $d_{dst}$ can be the same device. The set of all $N_P$ possible paths between $d_{src}$ and $d_{dst}$ can be represented by set $P=\{P_i \; | i\in[N_P]\}$. Each path is a sequence of devices $P_i=\{p_{i}^{j} \; | \; j\in[|P_i|], p_{i}^{j} \in D\}$, where $|P_i|$ is the length of path $P_i$.

There are four types of decision variables: 
(1) The deployment decision variable $X=(x_{ijk})_{T_s \times D_s \times N_D}$;
(2) The device usage variables $Y=(y_{i})_{1\times N_D}$, where each $y_i \in \{0, 1\}$;
(3) The path decision variable $Z=(z_{i})_{1\times N_P}$ to indicate which path is chosen; 
(4) The last stage deployment decision variable $C=({c_{ij})_{N_P \times N_D}}$. These variables are used to indicate where the ML pipeline ends, as packet sizes are smaller after the ML computation ends. All of these decision variables are in the range of $[0, 1]$, and a value of 1 indicates a decision at its corresponding position.

\subsection{Optimization Objectives}
\sysname{} aims at optimizing three objectives when generating an optimal deployment plan: (1) minimize inference serving time, (2) minimize the number of data planes used, and (3) minimize the overhead imposed on network traffic. 

\boldpara{Objective \#1: Minimize per-packet request serving time.} The per-packet request serving time is the duration between when an inference request is sent and when the response arrives at the destination. 
The per-packet request serving time is the sum of total transmission delay $\mathbb{J}_{trsmt}$ (pushing a packet on the wire), total propagation delay $\mathbb{J}_{prop}$ (packet propagating data planes), and total execution time $\mathbb{J}_{exe}$ (running inference) on one or more data planes. The packet transmission time per device or host is $l_t^r$ and $l_t^s$ for request and response packets, respectively. 
The propagation delay per hop is $l_p$. 
The total propagation delay $\mathbb{J}_{prop}$ depends on the length of the routing path, or the number of hops. We define the execution time at each switch $l_e$, as it is a constant number across all devices. The per-packet inference serving time is calculated as 

\begin{equation*}
\begin{aligned}
    \mathbb{J}_L &= \mathbb{J}_{exe} + \mathbb{J}_{prop} + \mathbb{J}_{trsmt} \\
                 &= \sum_{i=1}^{N_D} l_e \cdot y_i + \sum_{i=1}^{N_P} l_p \cdot z_{i} \cdot |P_i| \\
                 &+ \sum_{i=1}^{N_P}\sum_{j=1}^{N_D} (l_t^r \cdot pos(i, j) + l_t^s \cdot (|P_i| - pos(i,j))) \cdot c_{ij} \\
\end{aligned}
\end{equation*}
where $pos(i,j)$ is the position of device $d_j$ on path $P_i$. If $d_j$ is not on $P_i$, $pos(i,j)=0$.

\boldpara{Objective \#2: Minimize the number of programmable data planes.} 
The previous goal encourages the optimizer to use shorter paths, but a shorter path does not guarantee the minimum number of \textit{programmable} data planes in use. Using fewer programmable devices reduces overall resource usage. 
The number of used data planes can be written: $\mathbb{J}_{D}=\sum_{i=1}^{N_D} y_i$.

\boldpara{Objective \#3: Minimize network overhead.}
Performing in-network classification will inevitably create additional traffic and impose more overhead on the network. Even on the same path, different deployment strategies could result in different network overheads. \sysname{} considers this factor and tries to minimize the additional total number of bytes that must be added to a packet (\eg{} features, intermediate results). Assume that the packet sizes for \sysname{} request and response packets are $s_p^{rq}$ and $s_{p}^{rs}$ bytes. We calculate the overhead objective as:
\begin{equation*}
    \mathbb{J}_O = \sum_{i=1}^{N_P}\sum_{j=1}^{N_D} (s_{p}^{rq} \cdot pos(i, j) + (|P_i| - pos(i,j)) \cdot s_{p}^{rs}) \cdot c_{ij}
\end{equation*}

\smallskip
\noindent
\textbf{Final optimization goal.} The goal of the multi-objective optimization problem is to find a deployment strategy that jointly optimizes device usage, network overheads, and inference serving time at the same time. 
The global objective is:
\begin{equation*}
    \mathbb{J} = w_L \cdot \mathbb{J}_L + w_D \cdot \mathbb{J}_D + w_O \cdot \mathbb{J}_O, \;w_L+w_D+w_O = 1
\end{equation*}
where $w_L, w_D, w_O \in \mathbb{R^+}$ are the weights of inference serving time, device usage, and network overheads, respectively. 
In our evaluation, we use an even distribution of weights (\S ~\ref{sec:eval}). However, the weights are user-defined parameters that users may configure to control the importance of each objective. 

\subsection{Constraints}
We define five types of constraints in our optimizer to ensure that the final deployment plan is feasible. More details, including the formulas, are in Appendix \ref{appen:constraints}.

\boldpara{Integrity Constraints.} 
(1) Every program stage will be deployed once and only once on the network. For SVM, an extra constraint is needed to make sure all tables serving the same hyperplane will be placed on the same device. Moreover, (2) only one path must be chosen.

\boldpara{Guarantee constraints.} 
If at least one stage is deployed on device $d_i$, the decision variable for $d_i$ is set to 1. If at least one stage is deployed on any device on path $p_i$, the decision variable for $p_i$ is set to 1. 

\boldpara{Resource limitation.}
The number of stages deployed on one device should not exceed the maximum number of stages available on each device.
The number of table entries for each table must not exceed the available spaces on the corresponding tables on the target stage.

\boldpara{Stage dependency.} 
After distribution, the stages must be in the correct order. Suppose stage $s_{m}$ is deployed on stage $q$ of device $p_i^{n}$ on an arbitrary path $P_i$. Then, the dependency constraint requires that all predecessor stages of $s_{m}$ must be deployed before $s_m$, on devices $\{p_i^0, p_i^1, \cdots, p_i^n\}$.

\boldpara{Position of the last stage.} 
If a path $p_i$ is not chosen, then the last stage should not be deployed on path $p_i$. If a device $d_i$ is not chosen, then the last stage should not be deployed on that device. If the last stage is placed on device $d_k$ and on path $p_i$, then the corresponding decision variable, $c_{ij}$, must be 1.

\section{Runtime Programmability}
\label{sec:run_prog}

\sysname{} achieves runtime programmability through carefully designed control plane and data plane programs. 

\subsection{Data Plane}

\begin{table}
\centering
\resizebox{\columnwidth}{!}{
  \begin{tabular}{lccc}
    \toprule
    \textbf{Table Name} & \textbf{Matching Type} & \textbf{Arguments} & \textbf{Returns} \\
    \midrule
    dt\_layer & ternary & feature, prev\_status\_code, priority & status\_code \\
    dt\_predict & exact & status codes & dt classification result \\
    multitree\_voting & exact & all dt codes & voting result \\
    \midrule
    svm\_mul & exact & feature & product \\
    svm\_predict & exact & hyperplane codes & svm classification result \\
    \bottomrule
 \end{tabular}
}
\caption{Pre-defined set of tables for ML in \sysname{}.}
\label{tab:data_plane_tables}
\end{table}

In essence, \sysname{} trades \textit{generality} for \textit{optimized resource utilization}. By pre-defining specialized, ML-specific tables, \sysname{} achieves runtime programmability more effectively, supporting a large number of features and deeper tree structures.
In particular, we define five types of tables. Table~\ref{tab:data_plane_tables} shows the name, matching type, arguments, and the actions performed by these tables. \texttt{DT\_layer} and \texttt{DT\_predict} implement the inference procedure of decision trees, and \texttt{Multitree\_voting} performs the final classification step if the model is a multi-tree model. For SVM models, \texttt{SVM\_mul} tables holds precomputed multiplication results and \texttt{SVM\_predict} makes predictions.

\begin{figure*}[t]
  \centering
  \includegraphics[width=0.75\linewidth]{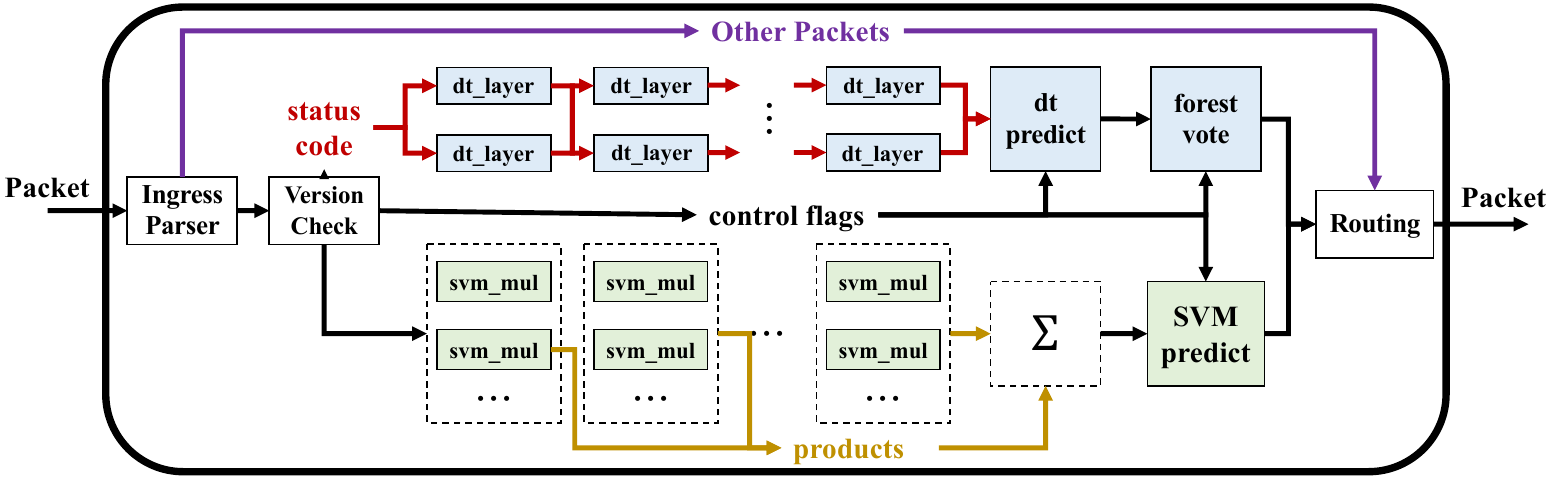}
  \caption{The data plane design of \sysname{}.}
  \label{fig:acorn_data_plane}
\end{figure*}

Figure~\ref{fig:acorn_data_plane} shows \sysname{}'s data plane program design.
The data plane program can perform pure forwarding/routing or ML inference before forwarding packets using its parser to identify different types of packets. 
The tables for ML processing are organized into two pipelines across multiple stages. Each pipeline is dedicated to processing a type of ML model, and at each stage, the tables associated with the same pipeline are grouped into blocks. Tables in the same block are independent and do not rely on each other's outputs. Blocks from different pipelines can be placed at the same stage, and they share the on-device resources such as TCAM and SRAM. For example, at stage 1 in Figure~\ref{fig:acorn_data_plane}, there is one \textbf{Tree Block} from the tree pipeline and one \textbf{SVM Block} from the SVM pipeline.

Currently, there are two types of pipelines: tree pipelines and SVM pipelines. Tree pipelines are made up of \textbf{DT\_layer}, \textbf{DT\_predict}, and \textbf{Multitree\_voting} tables. At each stage, two \textbf{DT\_layer} tables are grouped into a block. Each table accepts status codes from the previous layer as part of the keys, and updates new status codes for the next layers. For SVM pipelines, multiple \textbf{SVM\_mul} can be grouped into one block and placed at the same stage. These tables perform multiplications and store products into local variables, which are summed and used for predictions.

This data plane design conveys many benefits. First, \textbf{DT\_layer} tables mostly use TCAMs for ternary matching while \textbf{SVM\_mul} tables only use SRAMs. Placing them at the same stage maximizes the usage of on-device resources and saves pipeline stages for other tasks. Second, for each pipeline, placing multiple tables at one stage effectively increases the number of input features supported by \sysname{} at each switch, and thus the entire system. 
Third, tables are organized in a clear layered structure, making it easy to distribute, offload, and manage ML models on the data plane. 

\subsection{Control Plane}
 When a new task is submitted to \sysname{}, the control plane program sends control packets to all switches to the network to acquire resource usage information. For example, \sysname{}'s control plane estimates memory usage by acquiring the number of table entries installed on each switch, and stores this information locally for future use. Once the distribution plan and the table entries are generated, the control plane program updates table entries, and then informs the client that the network is ready to receive request packets. 

\section{Implementation and Evaluation}
\label{sec:eval}

In this section, we briefly describe our implementation, then present our experiment setup and evaluation results. 


\subsection{Implementation}

The control plane program is implemented with Python 3.13.7, comprising approximately 1200 lines of code. The ILP solver is implemented using the \textit{milp} function in the Python \textit{SciPy 1.10.1} package.
\sysname{} data plane program is implemented in P4\_16 with about 3000 lines of code. To communicate with the switches during runtime, we utilize the P4Runtime's \textit{gRPC} module. \sysname{} currently supports TNA and T2NA architecture. However, \sysname{} implementation is intentionally match-action table-centric, making it generic to any architecture or target. We discuss this more in Section~\ref{sec:discuss}.

\subsection{Experiment Setup}
\boldpara{Benchmarks.} We compare \sysname{} with state-of-the-art in-network ML systems: SwitchTree~\cite{lee2020switchtree}, LEO~\cite{jafri2024leo}, and DINC~\cite{zheng2023dinc}.
We excluded previous in-network ML systems that only support a single use case with hardcorded features since a fair comparison with various workloads is impossible with such systems (\eg{} \cite{akem2023flowrest, xie2022mousika, netbeacon, akem2024jewel}).

\begin{itemize}
    \item \textbf{SwitchTree~\cite{lee2020switchtree}} is one of the first attempts that embeds random forest models into the P4 data plane. Although SwitchTree uses flow-level features extracted from the data plane to perform only traffic analysis, it has been used as a benchmark in other works~\cite{jafri2024leo}, and the data plane design can be easily modified to support general use cases. Hence, we use it as one of the benchmark systems. We built a P4 program for SwitchTree by modifying its publicly available code~\cite{ST_code}.

    \item \textbf{LEO~\cite{jafri2024leo}} is a single-switch system that enables in-network ML traffic classification for decision tree models at line rate by proposing a new online model for decision trees, which claims to be more efficient and capable than IIsy~\cite{zheng2024iisy} and SwitchTree~\cite{lee2020switchtree}. We used the code generator in LEO's GitHub repository~\cite{LEO_code} and generated a data plane program that supports the largest number of features, and used that in our experiments.

    \item \textbf{DINC~\cite{zheng2023dinc}} is a distributed in-network computing system that distributes workloads to the network and performs packet-level inference. DINC uses Planter~\cite{zheng2021planter, zheng2024planter}, which is built upon IIsy~\cite{zheng2024iisy}, to represent ML models and optimizes for a distribution plan using an ILP solver. We implement the control plane logic, including the planner, of DINC using the source code available on GitHub~\cite{DINC_code}. We implemented the data plane code referring to the available code for Planter~\cite{Planter_code}.
\end{itemize}

\boldpara{12 machine learning workloads.} We evaluated our system using a variety of workloads based on works from different application domains, including intrusion detection systems (IDS), traffic classification (TC), video quality inference (VQI), and general machine learning (GML). Table~\ref{tab:workload} shows the list of works, datasets, and the model type.
We followed the data preprocessing steps in these works and built models that mimic the results posted in each paper. For datasets that are still beyond the capability of our current system, we used recursive feature elimination (RFE)~\cite{rfe} to reduce the number of features to the number listed in Table~\ref{tab:workload}. 
More details about the workloads are in Appendix~\ref{appen:datasets}.

\begin{table}[t]
\centering
\resizebox{\columnwidth}{!}{
  \begin{tabular}{c cccc}
    \toprule
    \textbf{Task} & \textbf{Work} & \textbf{Dataset} & \textbf{Model} \\
    \midrule
     \multirow{3}{*}{IDS} & \multirow{2}{*}{MultiTree~\cite{gao2019adaptive}}  &  \multirow{2}{*}{NSL-KDD~\cite{nsl-kdd}} & DT \circled{1} \\
     & & & SVM \circled{2} \\
     \cmidrule{2-4}
         & Jewel\cite{akem2024jewel} & UNSW-IoT~\cite{unsw-iot} & RF \circled{3}  \\
    \midrule
     \multirow{4}{*}{TC} &  \multirow{2}{*}{LEO~\cite{jafri2024leo}} & CIC-IDS2017~\cite{cicids2017} & DT \circled{4} \\
     & & UNSW-NB15~\cite{unswnb15} & DT \circled{5} \\
     \cmidrule{2-4}
     & Mousika~\cite{xie2022mousika} & ISCXVPN16~\cite{iscxvpn16} & RF \circled{6} \\   
     \cmidrule{2-4}
     & Planter~\cite{zheng2024planter} & CIC-IDS2017~\cite{cicids2017} & SVM \circled{7} \\
    \midrule
     VQI & VCAML~\cite{sharma2023estimating} & VCAML (Webex)~\cite{sharma2023estimating} & RF \circled{8} \\
    \midrule
     \multirow{5}{*}{GML} &  Planter~\cite{zheng2024planter} & Iris~\cite{iris_53} & SVM \circled{9} \\
     \cmidrule{2-4}
     & \multirow{2}{*}{Image Classification} & Handwritten Digits~\cite{sklearn_digits} & RF \circled{10}  \\
     &  & MNIST~\cite{mnist} & DT \circled{11}  \\
     \cmidrule{2-4}
     & Tabular Classification & SATDAP~\cite{student_performance} & DT \circled{12} \\
    \bottomrule
 \end{tabular}
}
\caption{Machine learning workloads used in the evaluation.}
\label{tab:workload}
\end{table}

\begin{table}[t]
\centering
\footnotesize
\begin{tabular}{c c c c c }
\toprule
\textbf{Model type} & \textbf{SwitchTree} & \textbf{LEO} & \textbf{DINC} & \textbf{\sysname{}} \\
\midrule
DT  & 16                          & 10                          & 40 & 46 \\
\midrule
RF  & \cellcolor[HTML]{C0C0C0}N/A & \cellcolor[HTML]{C0C0C0}N/A & 20 & 46 \\
\midrule
SVM & \cellcolor[HTML]{C0C0C0}N/A & \cellcolor[HTML]{C0C0C0}N/A & \purple{8}  & \purple{8}  \\
\bottomrule
\end{tabular}
\caption{The maximum number of features for per model type supported by each evaluated system.}
\label{tab:maxfeatures}
\end{table}

\boldpara{\textit{Note}:}
During our experiments, We found that \sysname{}'s SVM operation was not summing integers correctly. Through our interaction with Intel, it has been confirmed that this is due to a \textit{compiler bug}, and it is not an implementation issue. Thus, we had to use a reduced pipeline for SVMs that supports only up to 8 features.
To compensate for this, we wrote a simulator that simulates the correct computation process of our SVM pipeline, which supports 46 features. The results are reported in a separate column in Table~\ref{tab:acc_sim_svm}. Details about the simulator are mentioned in Appendix~\ref{appen:exp_setups}. 

\boldpara{Physical testbed.} Our testbed has two Tofino 2 100G switches and two HPE ProLiant DL385 Gen11 servers. We use one server as the source machine, and the other server as the destination machine. 
For experiments that require more than two switches, we utilize four separate pipelines in each switch, making 
packets travel between up to eight pipelines. 

\subsection{Classification Results}
\label{sec:classification}
We first report on the classification performance of \sysname{} and the benchmark systems in ML workloads. We evaluate the performance using the following metrics: (1) accuracy (Acc), or the accuracy of classifications; (2) f1 score (Macro-F1), or the macro F1 score of classifications; (3) Cohen's Kappa (K)~\cite{cohen1960kappa}, to describe the level of agreement between the predictions of in-network systems and the server-based system. Cohen's Kappa metric assesses the inter-rater reliability for qualitative (categorical) items. Unlike simple percent agreement, it adjusts for chance agreement, making it a more robust measure---particularly in the presence of class imbalance.
We also reported the number of features used in each setup, as the same setup does not apply to all systems. 

\boldpara{Result.} We present the results for tree-based models in Table~\ref{tab:acc_sim}. For most workloads, \sysname{} achieves better accuracy and macro-F1 scores compared to other benchmark systems, which is the benefit of supporting more features. The Cohen's Kappa scores indicate that for most workloads, the impact of imprecision caused by model translation is minimal, and the prediction results are trustworthy. There are some cases where benchmark systems achieve better performance than \sysname{}. For example, for workload \circled{12}, SwitchTree achieves the best result with 16 features, better than DINC with 36 features and \sysname{} with 24 features. These cases indicate that more features do not guarantee better performance.
However, note that \sysname{} can also deploy the best model with less features for each task in the table, demonstrating its high expressiveness.

\boldpara{Analysis.}
DINC~\cite{zheng2023dinc} uses Planter~\cite{zheng2024planter} to represent ML models by mapping features to encoded ranges and makes decisions by listing all possible combinations of feature codes. The increase in the number of features and the tree depth eventually results in a factorial-like growth of entries in the decision table, making the model unable to deploy. DINC compensates for that by using more devices. In each device, DINC minimizes packet size by micro-managing the bits and packing multiple variables into one field. However, these operations on bits consume a lot of ALUs and are toxic to the P4 compiler, which easily leads to compiler errors. Thus, in datasets with multiple small classes, for example, UNSW-IoT and MNIST, this forces models to underfit, causing low accuracy and K values.

We also measure the performance for SVMs and list the results in Table~\ref{tab:acc_sim_svm}. For all workloads, both DINC and \sysname{} achieve similar and close to CPU results. In terms of accuracy and Cohen's Kappa metric, both systems are on the same level. Our simulator uses more features to reflect the performance of our original design. Results show that our original SVM pipeline design makes predictions that align with CPU results, but using more features do not always mean better accuracy.


\begin{table*}
\centering
\resizebox{\textwidth}{!}{
  \begin{tabular}{c cccc cccc cccc cccc cccc}
    \toprule
    \multirow{2}{*}{\textbf{Workload}} & \multicolumn{4}{c}{\textbf{SwitchTree}} & \multicolumn{4}{c}{\textbf{LEO}} & \multicolumn{4}{c}{\textbf{DINC}} & \multicolumn{4}{c}{\textbf{\sysname{}}} \\
    \cmidrule(lr){2-5}\cmidrule(lr){6-9}\cmidrule(lr){10-13}\cmidrule(lr){14-17}
    & \textbf{Acc} & \textbf{Macro-F1} & \textbf{K} & \textbf{Feat} & \textbf{Acc} & \textbf{Macro-F1} & \textbf{K} & \textbf{Feat} & \textbf{Acc} & \textbf{Macro-F1} & \textbf{K} & \textbf{Feat} & \textbf{Acc} & \textbf{Macro-F1} & \textbf{K} & \textbf{Feat} \\
    \midrule
     \circled{1} & 0.810 & 0.810 & 0.822 & 16 & 0.808 & 0.807 & 0.999 & 10 & \textbf{0.859} & \textbf{0.852} & 0.795 & 32 & 0.823 & 0.823 & \textbf{1.0} & 46 \\
     \midrule
      \circled{3} & \cellcolor[HTML]{C0C0C0}N/A & \cellcolor[HTML]{C0C0C0}N/A & \cellcolor[HTML]{C0C0C0}N/A & \cellcolor[HTML]{C0C0C0}N/A & \cellcolor[HTML]{C0C0C0}N/A & \cellcolor[HTML]{C0C0C0}N/A & \cellcolor[HTML]{C0C0C0}N/A & \cellcolor[HTML]{C0C0C0}N/A  & 0.330 & 0.190 & 0.293 & 20 & \textbf{0.841} & \textbf{0.842} & \textbf{0.923} & 24 \\
     \midrule
      \circled{4} & 0.994 & 0.994 & 0.989 & 16 & 0.999 & 0.999 & 0.999  & 10 & 0.997 & 0.997 & 0.995 & 40 & \textbf{0.999} & \textbf{0.999} & \textbf{0.999} & 46 \\
      \midrule
       \circled{5} & 0.829 & 0.823 & 0.996 & 16 & 0.852 & 0.849 & 0.999 & 10 & 0.790 & 0.778 & 0.754 & 32& \textbf{0.871} & \textbf{0.869} & 0.993 & 26\\
      \midrule
       \circled{6} & \cellcolor[HTML]{C0C0C0}N/A & \cellcolor[HTML]{C0C0C0}N/A & \cellcolor[HTML]{C0C0C0}N/A & \cellcolor[HTML]{C0C0C0}N/A & \cellcolor[HTML]{C0C0C0}N/A & \cellcolor[HTML]{C0C0C0}N/A & \cellcolor[HTML]{C0C0C0}N/A & \cellcolor[HTML]{C0C0C0}N/A & 0.749 & 0.416 & 0.525 & 20 & \textbf{0.897} & \textbf{0.709} & 0.994 & 16 \\
      \midrule
       \circled{8} & \cellcolor[HTML]{C0C0C0}N/A & \cellcolor[HTML]{C0C0C0}N/A & \cellcolor[HTML]{C0C0C0}N/A & \cellcolor[HTML]{C0C0C0}N/A & \cellcolor[HTML]{C0C0C0}N/A & \cellcolor[HTML]{C0C0C0}N/A & \cellcolor[HTML]{C0C0C0}N/A & \cellcolor[HTML]{C0C0C0}N/A & 0.997 & 0.906 & 0.812 & 14 & \textbf{0.999} & \textbf{0.970} & \textbf{1.0} & 14 \\
      \midrule
       \circled{10} & \cellcolor[HTML]{C0C0C0}N/A & \cellcolor[HTML]{C0C0C0}N/A & \cellcolor[HTML]{C0C0C0}N/A & \cellcolor[HTML]{C0C0C0}N/A & \cellcolor[HTML]{C0C0C0}N/A & \cellcolor[HTML]{C0C0C0}N/A & \cellcolor[HTML]{C0C0C0}N/A & \cellcolor[HTML]{C0C0C0}N/A & 0.711& 0.686 & 0.657& 20 & \textbf{0.886} & \textbf{0.887} & \textbf{1.0} & 38\\
      \midrule
       \circled{11} & 0.649 & 0.639 & 0.764 & 16 & 0.686 & 0.679 & 0.999 & 10 & 0.593& 0.589 & 0.546 & 32 & \textbf{0.823} & \textbf{0.821} & 0.999 & 46\\
      \midrule
       \circled{12} & \textbf{0.767} & \textbf{0.704} & 0.897 & 16 & 0.716 & 0.638 & 1.0 & 10 & 0.749 & 0.653 & 0.567 & 36 & 0.707 & 0.657 & 1.0 & 24 \\
    \bottomrule
 \end{tabular}
}
\caption{Tree-based model workloads on \sysname{} vs. benchmarks.}
\label{tab:acc_sim}
\end{table*}

\begin{table}[h]
\centering
\resizebox{\columnwidth}{!}{
  \begin{tabular}{c cccc cccc cccc cccc}
    \toprule
    \multirow{2}{*}{\textbf{Workload}} & \multicolumn{4}{c}{\textbf{DINC}} & \multicolumn{4}{c}{\textbf{\sysname{} (Simulator)}} & \multicolumn{4}{c}{\textbf{\sysname{}}} \\
    \cmidrule(lr){2-5}\cmidrule(lr){6-9}\cmidrule(lr){10-13}
    & \textbf{Acc} & \textbf{W-F1} & \textbf{K} & \textbf{Feat} & \textbf{Acc} & \textbf{W-F1} & \textbf{K} & \textbf{Feat} & \textbf{Acc} & \textbf{W-F1} & \textbf{K} & \textbf{Feat} \\
    \midrule
     \circled{2} & 0.751 & 0.750 & {0.804} & 8 & 0.749 & 0.748 & 0.999 & 46 & \textbf{0.798} & \textbf{0.798} & \textbf{0.999} & 5 \\
     \midrule
       \circled{7} & 0.997 & 0.997 & 0.994 & 8 & 0.999 & 0.999 & 1.0 & 46 & \textbf{0.999} & \textbf{0.999} & \textbf{0.999} & 8 \\
      \midrule
       \circled{9} & 0.97 & 0.97 & 1.0 & 4 & 0.933 & 0.933 & 1.0 & 4 & \textbf{0.933} & \textbf{0.933} & \textbf{1.0} & 4 \\
    \bottomrule
 \end{tabular}
}
\caption{SVM workloads on \sysname{} vs. benchmarks.}
\label{tab:acc_sim_svm}
\end{table}

\subsection{Request serving latency}

\textit{Request serving latency} is the total delay between when a request is sent and the result is received. 
As in Section~\ref{sec:opt}, this includes the transmission delay, propagation delay, and execution time.
We measured the per-packet request serving time to evaluate the speed and stability of \sysname{} comparing \sysname{} with a server-based system.
We collect two timestamps at a single host: TX and RX timestamps. We use the same host machine for sending the request and receiving the result to use the same clock without time synchronization issues.
The request serving time is $t_{s}=RX-TX$. 
 
For server-based solution, we fix the number of hops to six, assuming the request client machine is in one rack and the server, where the ML model runs, is in another rack, thus two machines are connected through two top-of-rack switches.  For \sysname{}, the number of hops varies, influenced by the ML model size and workload. The final number of hops is determined by \sysname{}'s deployment optimizer. 

\begin{figure}[t]
    \centering
    \begin{subfigure}[b]{0.49\columnwidth}
        \includegraphics[width=\linewidth]{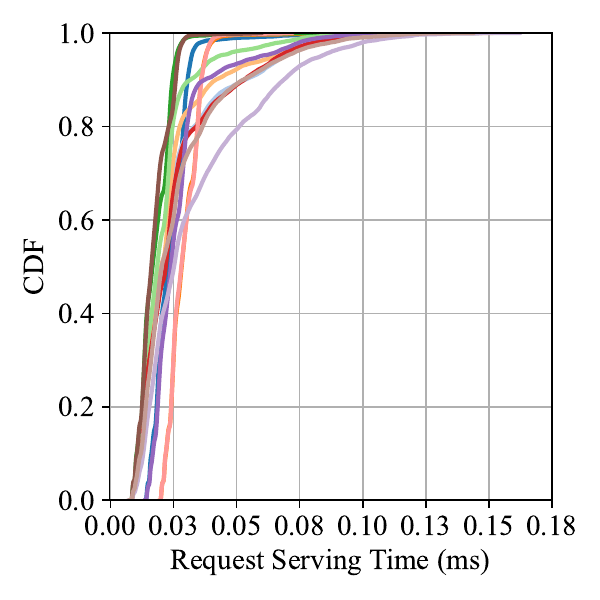}
        \caption{\sysname{}}
        \label{fig:inference_time:switch}
    \end{subfigure}
    \hfill
    \begin{subfigure}[b]{0.49\columnwidth}
        \includegraphics[width=\linewidth]{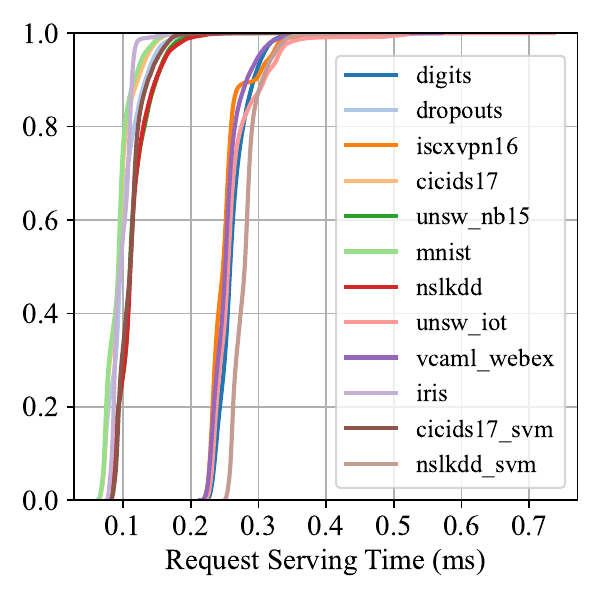}
        \caption{Server}
        \label{fig:inference_time:server}
    \end{subfigure}
    \caption{Serving time for the workloads: \sysname{} vs. server}
    \label{fig:inference_time}
\end{figure}

Figure~\ref{fig:inference_time} shows our result. In general, \sysname{} serves requests faster than server-based systems by 65--90\%. 
For most workloads, requests are served within 0.12 ms on \sysname{} in consistent intervals, with very few outlier cases. On servers, we notice that the request serving time is affected by the model type. For decision tree models and SVMs, the majority of requests can be completed in 0.3 milliseconds. However, for random forest models, the request serving time doubles. 

\begin{figure}[t]
    \centering
    \includegraphics[width=\linewidth]{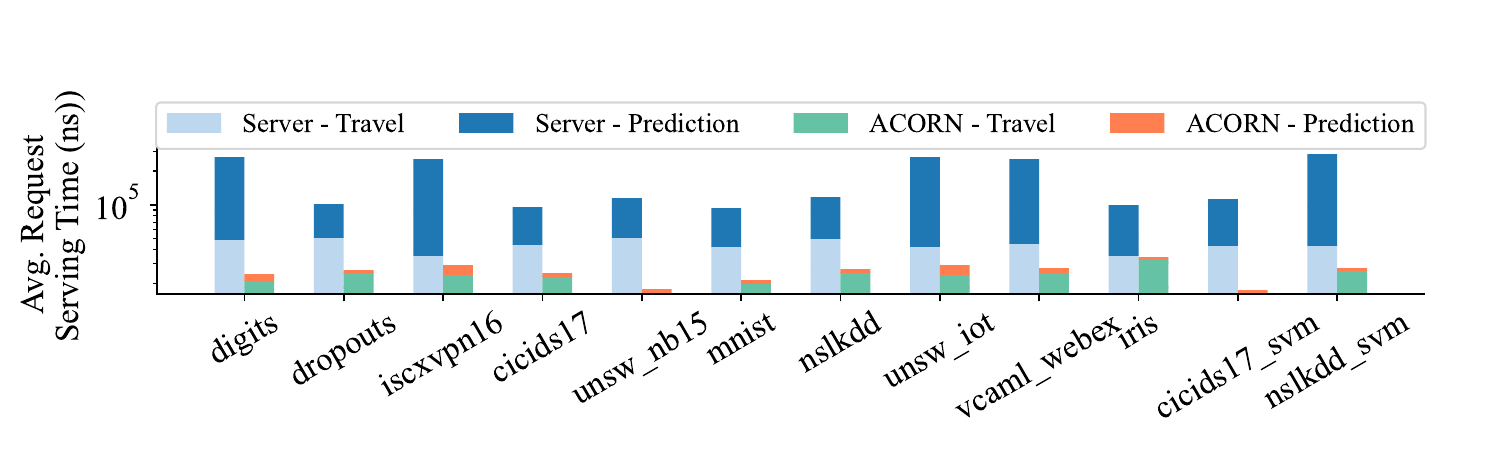}
    \caption{Average request serving time in logscale.}
    \label{fig:avg_serving_time}
\end{figure}

Since the total latency is a combination of different factor, we break it down by function. 
We define \textit{prediction latency} as the latency incurred purely on running an inference task on data planes or a server, while \textit{travel latency} is for the rest, which is dominated by transmission and propagation delay in the network. 
Figure~\ref{fig:avg_serving_time} shows, on average, how \sysname{} and server-based approach compare. 
\sysname{} provides a speedup of 98x in \textit{prediction latency}.
Although servers are equipped with powerful processing units, we still notice that in most cases, the delay is caused by making ML predictions on the server. 
\sysname{} also provides an opportunity to shorten the \textit{traveling latency} since a packet does not have to visit another server's host network stack, which may add significant latency~\cite{cai2021understanding,agarwal2023host,vuppalapati2024understanding}.
For server-based approach, running the ML model in a server closer to the client machine or even on the same machine may eliminate the \textit{traveling latency}. Yet, even without the \textit{traveling latency} in the server-based approach, \sysname{} demonstrates shorter overall latency.

\subsection{Optimization}
To measure \sysname{} optimizer's performance and how it scales, we evaluate the optimizer on four simulated datacenter topologies with varying sizes: K-ary Fat-Tree~\cite{fat_tree}, DCell~\cite{dcell}, BCube~\cite{bcube}, and Jellyfish~\cite{jellyfish}. In Table~\ref{tab:topo}, for Fat-Tree, \texttt{k} is the number of pods; for DCells, \texttt{n} and \texttt{k} are the number of sub-cells and the number of layers; \texttt{n} and \texttt{k} in BCubes represent the number of ports in each switch, and the number of levels; Jellyfish topologies are defined by \texttt{n}, the number of switches and \texttt{d}, the degree of each node.
The source and destination nodes are two different servers in the same network. 
We selected the parameters so that there are a similar number of switches in the network across different topologies. We use respective routing rules and paths for each type of topology.

\begin{table}[t]
\centering
\resizebox{\columnwidth}{!}{
  \begin{tabular}{c cccc}
    \toprule
      & \textbf{Fat-Tree (k)} & \textbf{DCell (n, k)} & \textbf{BCube (n, k)} & \textbf{Jellyfish (n, d)}  \\
    \midrule
    Setup 1 & 12 & (3, 2) & (5, 2) & (80, 3) \\
    \midrule
    Setup 2 & 16 & (4, 2) & (7, 2) & (125, 4) \\
    \midrule
    Setup 3 & 20 & (5, 2) & (8, 2) & (170, 3) \\
    \bottomrule
 \end{tabular}
}
\caption{Information about the used topology.}
\label{tab:topo}
\end{table}

\begin{figure}[t]
    \centering
    \begin{subfigure}[b]{0.43\linewidth}
        \includegraphics[width=\linewidth]{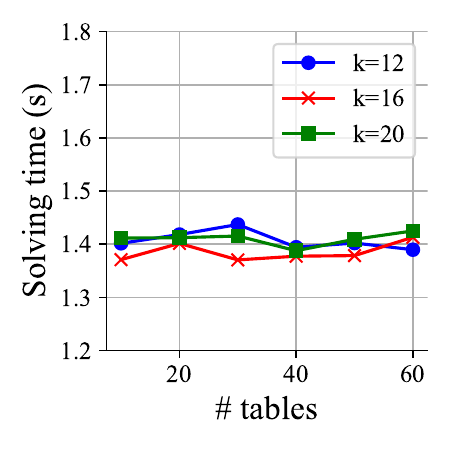}
        \caption{Fat-tree}
        \label{fig:topo:fattree}
    \end{subfigure}
    \hfill
    \begin{subfigure}[b]{0.43\linewidth}
        \includegraphics[width=\linewidth]{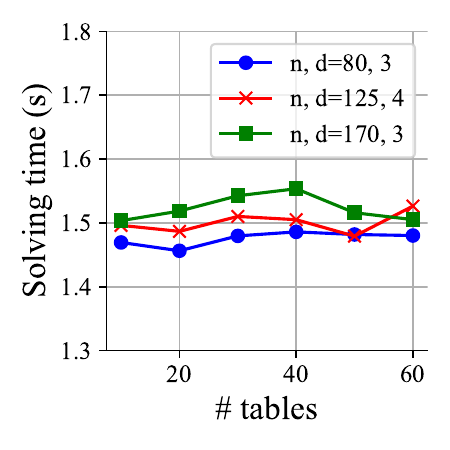}
        \caption{Jellyfish}
        \label{fig:topo:jellyfish}
    \end{subfigure}
    \hfill
    \begin{subfigure}[b]{0.43\linewidth}
        \includegraphics[width=\linewidth]{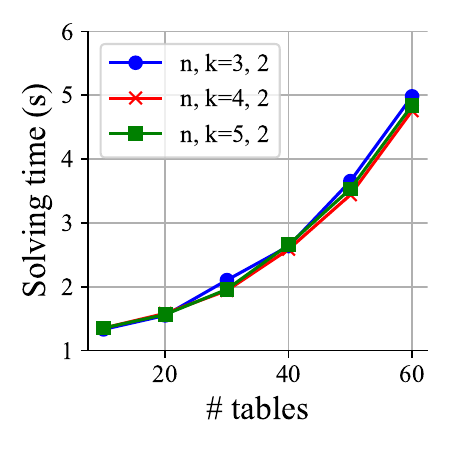}
        \caption{DCell}
        \label{fig:topo:dcell}
    \end{subfigure}
    \hfill
    \begin{subfigure}[b]{0.43\linewidth}
        \includegraphics[width=\linewidth]{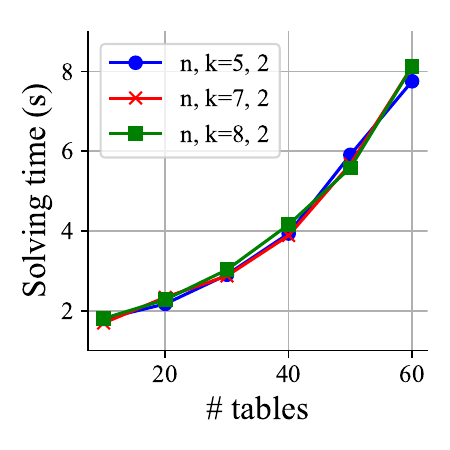}
        \caption{BCube}
        \label{fig:topo:bcube}
    \end{subfigure}
    \caption{Solving time of optimizer for network topologies.}
    \label{fig:scalability}
\end{figure}

We present the results in Figure~\ref{fig:scalability}. For all cases, our optimizer solves the problems within 10 seconds. 
We observe that for Fat-tree and Jellyfish topologies, the optimization time remains steady regardless of the topology scale or the program size. For BCube and DCell, the optimization time increases as the program size increases, because the number of decision variables $x_{i,j,k}$ increases as the program size increases. 
However, the size of the graph does not affect the optimization time. Our parallelization optimization strategy decomposes the problem into sub-problems. Each subproblem only considers the devices on the path, which is a relatively small number. For example, \texttt{Jellyfish(125, 4)} increases the problem size by a multiplier of 45 compared to \texttt{Jellyfish(80, 3)}.
Yet, each sub-problem for \texttt{Jellyfish(125, 4)} only increases the problem size by a multiplier of 2. This increase is small for the optimizer, having negligible effect. 

\subsection{\sysname{} Resource Usage and Overhead}

We measure \sysname{}'s resource usage and how much overhead it incurs when running on a network. 

\boldpara{Static resource overhead.}
We measure the static resource usage on the data plane by analyzing the compilation log and conducting an ablation study for each component of the data plane program.
We use an Intel Tofino 1 switch as a reference\footnote{The ratio is smaller for a Tofino 2 switch, which has more resources.}.
Table~\ref{tab:res_overhead} shows the resource usage in percentages. In general, \sysname{}'s data plane imposes little static overhead on network devices.
For Packet Header Vector (PHV), PHV words store packet and metadata fields, and the relatively moderate usage of PHVs is inevitable, as the features are stored and transmitted in packet headers. The maximum usage of PHV for the basic program and all DT\_layer tables is 20\% of the total PHVs, which is acceptable.
Very Long Instruction Words (VLIWs) are needed to perform actions on packet header fields and local variables. Each pair of DT\_layer tables contain actions for all 46 features, and actions to modify local variables that store status codes and feature values. Thus, the VLIWs usage is also relatively high. Reducing the number of supported features in the data plane reduces the VLIW usage.

\begin{table}[t]
\centering
\resizebox{\columnwidth}{!}{
  \begin{tabular}{lcp{1cm}p{1cm}p{1cm}cccc}
    \toprule
    \textbf{Component} & \textbf{PHV} & \textbf{Exact Match Xbar} & \textbf{Ternary Match Xbar} & \textbf{Hash Bits} & \textbf{SRAM} & \textbf{TCAM} & \textbf{VLIW} \\
    \midrule
    Base program & 7.86 & 0.58 & 0 & 0.42 & 0.31 & 0 & 2.6 \\
    \midrule
    DT\_layer (2) & 12.14 (total) & 0 & 1.52 & 0 & 0.21 & 2.08 & 6.5 \\
    DT\_predict & 0.4 & 1.04 & 0 & 0.8 & 1.25 & 0 & 0.5 \\
    Multitree\_voting & 0 & 0.2 & 0 & 1.2 & 0.41 & 0 & 0.26 \\
    \midrule
    SVM\_mul & 0.17 & 0.13 & 0 & 0.8 & 0.41 & 0 & 0.13 \\
    SVM\_predict & 1.07 & 0.13 & 0 & 0.2 & 0.11 & 0 & 0.52 \\
    \bottomrule
 \end{tabular}
}
\caption{Resources usage (in \%) of \sysname{} data plane.}
\label{tab:res_overhead}
\end{table}

\boldpara{Pipeline stage usage.} 
We first study whether supporting tree-based models with more features would require more pipeline stages.
We do not study SVM models as they are not sensitive to stages. Table~\ref{tab:stage_usage} presents the usage of pipeline stages for each workload with respect to different numbers of features. For random forest models, we present the average stage usage over all trees in the model. 
We do not observe an obvious increasing trend in pipeline stage usage for any workload, which means supporting more features does not require more stages as a trade-off. Compared to 5-feature cases, some 45-feature cases require an extra 1--5 stages, which is an increase of 4--14\% compared to a 9x increase in feature numbers. In some cases, such as CICIDS17, Digits, ISCXVPN16, and MNIST, we even see fewer stages used for 45-feature cases.
Further investigation revealed that when some features are missing, the model has to use more layers to split the data samples, resulting in fewer features using more pipeline stages.

\begin{table}[t]
\centering
\resizebox{\columnwidth}{!}{
  \begin{tabular}{l ccccccccc}
    \toprule
    \multirow{2}{*}{\textbf{Workload}}  & \multicolumn{9}{c}{\textbf{Features}}  \\
    \cmidrule{2-10}
      & 5 & 10 & 15 & 20 & 25 & 30 & 35 & 40 & 45 \\
    \midrule
    CICIDS17 & 14 & 12 & 12 & 12 & 12 & 12 & 12 & 12 & 12\\
    \midrule
    Digits & 16.5 & 14.5 & 14.5 & 13.25 & 13.75 & 13.25 & 13.5 & 13.25 & 13.5 \\
    \midrule
    Dropouts & 24 & 21 & 21 & 22 & 23 & 25 & 25 & \cellcolor[HTML]{C0C0C0}N/A & \cellcolor[HTML]{C0C0C0}N/A \\
    \midrule
    ISCXVPN16 & 17.25 & 19.75 & 17.75 & 16.75 & \cellcolor[HTML]{C0C0C0}N/A & \cellcolor[HTML]{C0C0C0}N/A & \cellcolor[HTML]{C0C0C0}N/A & \cellcolor[HTML]{C0C0C0}N/A & \cellcolor[HTML]{C0C0C0}N/A \\
    \midrule
    MNIST & 33 & 30 & 26 & 26 & 28 & 26 & 24 & 25 & 30 \\
    \midrule
    NSL-KDD & 21 & 25 & 23 & 21 & 23 & 22 & 24 & 21 & 24 \\
    \midrule
    UNSW\_IoT & 35 & 37.75 & 39.5 & 38.25 & 36.75 & 38 & \cellcolor[HTML]{C0C0C0}N/A & \cellcolor[HTML]{C0C0C0}N/A & \cellcolor[HTML]{C0C0C0}N/A \\
     \midrule
    UNSW\_NB15 & 50 & 42 & 46 & 43 & 52 & 52 & 54 & 52 & 55 \\
     \midrule
    VCAML\_Webex & 7.25 & 8.5 & 7.25 & \cellcolor[HTML]{C0C0C0}N/A & \cellcolor[HTML]{C0C0C0}N/A & \cellcolor[HTML]{C0C0C0}N/A & \cellcolor[HTML]{C0C0C0}N/A & \cellcolor[HTML]{C0C0C0}N/A & \cellcolor[HTML]{C0C0C0}N/A  \\
    \bottomrule
 \end{tabular}
}
\caption{Pipeline stage usage of \sysname{}.}
\label{tab:stage_usage}
\end{table}

\begin{figure*}[t]
    \centering
    \begin{subfigure}[b]{0.24\textwidth}
        \includegraphics[width=\linewidth]{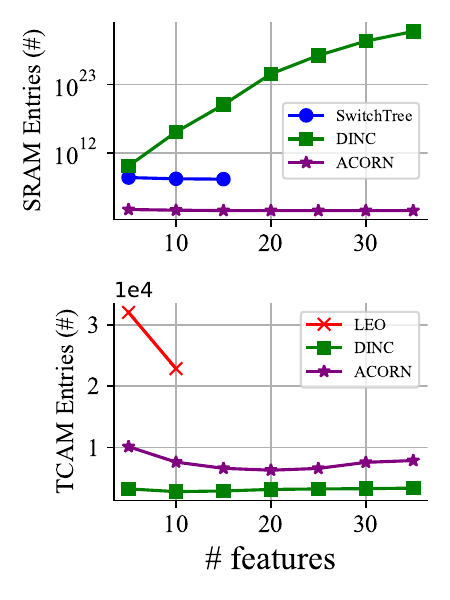}
        \caption{Dropouts}
        \label{fig:dyn_res:workload1}
    \end{subfigure}
    \hfill
    \begin{subfigure}[b]{0.24\textwidth}
        \includegraphics[width=\linewidth]{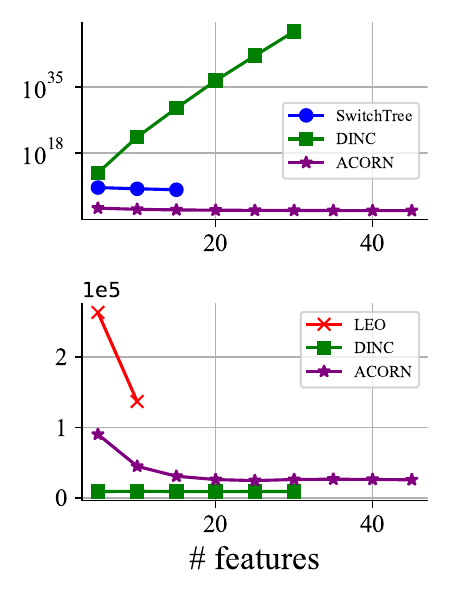}
        \caption{MNIST}
        \label{fig:dyn_res:workload2}
    \end{subfigure}
    \hfill
    \begin{subfigure}[b]{0.24\textwidth}
        \includegraphics[width=\linewidth]{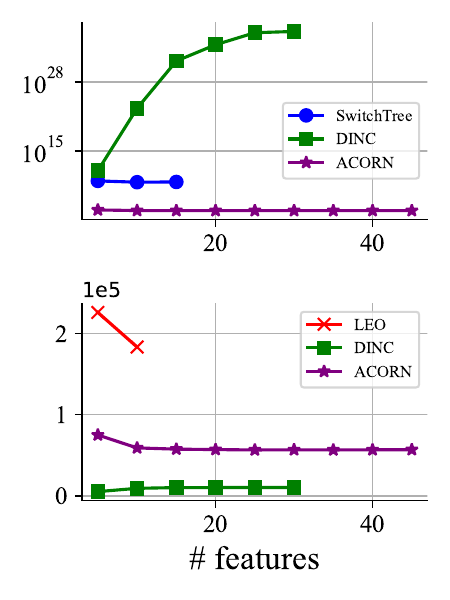}
        \caption{UNSW-NB15}
        \label{fig:dyn_res:workload3}
    \end{subfigure}
    \hfill
    \begin{subfigure}[b]{0.24\textwidth}
        \includegraphics[width=\linewidth]{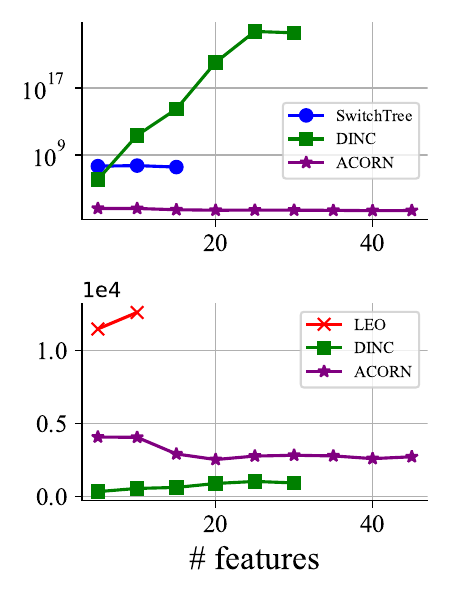}
        \caption{NSL-KDD (DT)}
        \label{fig:dyn_res:workload4}
    \end{subfigure}
    \caption{TCAM and SRAM usage of benchmark systems for tree models.}
    \label{fig:dyn_res}
\end{figure*}

\boldpara{SRAM and TCAM usage.}
To study the memory efficiency of \sysname{}, we vary the number of features and translate our workloads into table entries using \sysname{} and the benchmark systems. We measure the number of entries for SRAM and TCAM as the metric.
Figure~\ref{fig:dyn_res} shows our result. In all four workloads, DINC produces the fewest TCAM entries, followed by \sysname{} and LEO. Unlike \sysname{} and LEO, which use ternary matching to match on features at each stage, DINC uses TCAMs to match features to codes in one shot, which naturally uses fewer TCAM entries. 
\sysname{} still shows resource efficiency by using 2--3x fewer TCAM entries than LEO. 
We also observe an interesting trend that as the number of features increases, the TCAM usage of \sysname{} decreases in all four cases. As we mentioned above, with less features, tree models must use more layers and, as a result, more nodes to fit the datasets. The number of nodes is directly related to the number of TCAM entries for \sysname{}. 
This shows that supporting more features may also introduce resource benefits.
LEO also demonstrates this feature. However, in LEO, each table matches three inputs, and the combination of inputs increases the entry usage and offsets the benefits (Figure~\ref{fig:dyn_res:workload4}).

\sysname{} also requires fewer SRAM entries than SwitchTree and DINC. SwitchTree uses direct table lookups at each layer. Its SRAM usage is related to the precision of the inputs and the total number of nodes in the tree. 
DINC's decision table employs pure exact matching, and as our analysis in \S \ref{sec:classification}, the number of entries in the decision table increases factorially and rapidly reaches computationally infeasible numbers.
Meanwhile, the number of SRAM entries in \sysname{} is decided by the number of leaf nodes in the trees.

\boldpara{Network overhead.} 
In contrast to a run-to-completion (RTC) model, TNA and TNA2 architecture follows a pipeline model---if the program fits, packet processing happens at line rate. Yet, adding complex packet processing logic to the data plane still marginally impacts the packet processing rate. To measure how much latency \sysname{} would add when compared to a simple switching program, we generate a packet stream using 
\texttt{tcpreplay} with DPDK to constantly send jumbo frames with \sysname{} headers to the network to fill up the 10Gbps link. With different data plane programs loaded, we measure the latency and throughput to capture the impact and overhead of using \sysname{}.

\begin{figure}[t]
  \centering
  \includegraphics[width=\linewidth]{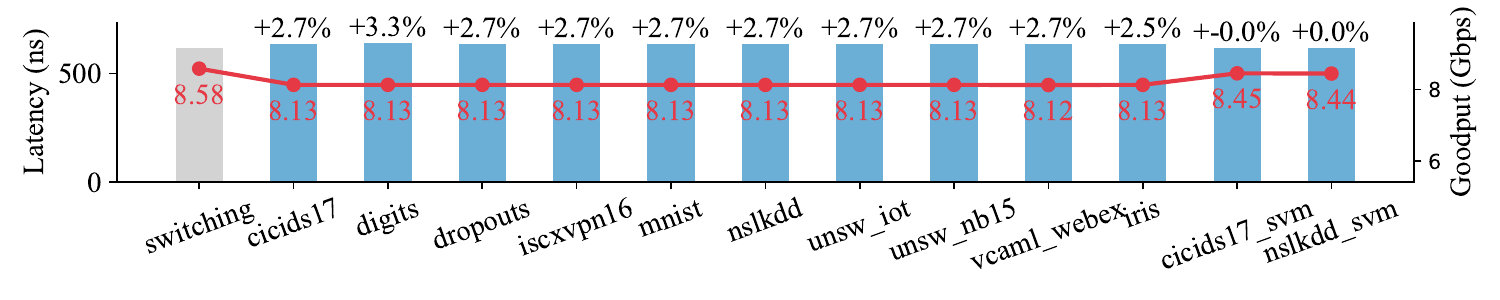}
  \caption{Latency and throughput overhead of \sysname{}.}
  \label{fig:latency_throughput}
\end{figure}

The average latency and goodputs are reported in Figure~\ref{fig:latency_throughput}. Compared to the statistics of pure switching for arbitrary traffic, \sysname{} only imposes an extra latency of 2.7--3.3\% to the data plane. \sysname{} also achieves near-line rate throughput for all workloads, with an impact of about 5\%. The result also shows consistent latency and throughput, demonstrating \sysname{}'s steady performance.
We do not expect much of a difference with a higher bandwidth or sending rate.

\section{Related Works}

\boldpara{In-network classification.} Most in-network classification works focus on traffic analysis. They extract features from packet metadata either at flow-level~\cite{busse2019pforest, xie2022mousika, zhang2021pheavy, akem2023flowrest, lee2020switchtree}, or at both packet and flow levels~\cite{netbeacon, akem2024jewel}. Other works, such as IIsy~\cite{xiong2019switches, zheng2024iisy}, Planter~\cite{zheng2021planter, zheng2024planter}, and LEO~\cite{jafri2024leo}, can perform \textit{general} ML classification tasks.
However, they are not capable of running large ML workloads with more than 40 features.

\boldpara{Neural networks.}
Several works successfully offload neural networks, such as binary neural networks~\cite{sanvito2018can, siracusano2018network, zheng2024planter} and deep neural networks~\cite{yan2024brain, zhang2025pegasus, paolini2024line}. 
However, the size of DNN models are inherently limited by the data plane's capability. To address this, 
several works leverage external hardware (\eg{} FPGAs~\cite{swamy2022taurus}, photonic SmartNIC~\cite{zhong2023lightning}).
Our work focuses on utilizing commodity line-rate P4-programmable data planes.

\boldpara{Distributed in-network computing.}
Researchers have been exploring approaches that allow programs to be deployed on multiple switches. 
Current distributed in-network computing systems involve transforming data plane programs into sub-pieces in the form of intermediate representations (\eg{} SPEED~\cite{chen2020speed}, Lyra~\cite{gao2020lyra}, Hermes~\cite{chen2022toward}, ClickINC~\cite{xu2023clickinc}), sub-programs (\eg{} FlightPlan~\cite{sultana2021flightplan}, DINC~\cite{zheng2023dinc}), or sub-models (\eg{} DUNE~\cite{butun2025dune}), and then distributing them to the network based on plans generated by an optimizer. Among these systems, only DINC~\cite{zheng2023dinc} and DUNE~\cite{butun2025dune} support ML programs, but neither supports runtime programmability, and only applies to a small set of applications and datasets.
~\cite{zhu2024conextsw} proposes a runtime-programmable in-network ML system, but does not provide any detailed design, implementation, or evaluation.

\noindent
\textbf{Runtime-configurable switches.} Our work achieves runtime programmability by dynamically managing on-device resources. Previous systems adopting similar idea either focus on specific tasks~\cite{zhou2020newton, zheng2022flymon}, or achieve runtime programmability for general tasks by using custom primitives to describe programs~\cite{das2023memory, khashab2024multitenant, yang2024p4runpro} and compile these programs to P4 using customized compilers. 
However, no previous effort were made to leverage these systems for in-network classification tasks, and they are not suitable for specific tasks such as ML.

\section{Discussion and Limitations}
\label{sec:discuss}

\boldpara{Hardware portability.} 
Due to hardware availability, we only evaluated \sysname{} on the Tofino architecture. However, our system relies solely on match-action tables, and does not rely on registers, or any Tofino-specific features such as packet circulation and packet mirroring. 
We expect that our system could be deployed on any other programmable data planes that support P4 and match-action tables (\eg{} Xsight Labs~\cite{www-xsight}, Cisco Silicon One~\cite{www-ciscosiliconone}, AMD Pensando~\cite{www-amdpensando}, NVIDIA BlueField~\cite{www-nvidiabluefield}, Silicom ThunderFjord~\cite{www-thunderfjord}, Intel IPU~\cite{www-intelipu}).

\boldpara{Multi-tenancy and user isolation.} 
\sysname{} supports multi-tenancy and user isolation because the data plane matches the tuple of (model\_id, version\_id) to distinguish between model types and different instances of the same model type. 
While we leave empirical evaluation for future work, our optimizer can already compute deployment plans under multi-tenant constraints. Despite the capability, we leave the experiments and evaluation for future work. 

\boldpara{Handling network failures and formal verification.} 
In this paper, we assumed that the network is absolutely reliable without any network failures, including link and switch failures, as well as packet loss. We recognize that this is an ideal assumption. Moreover, exploring fault tolerance and formal verification for in-network ML inference remains an open research direction. 
We leave this as future work. 

\boldpara{Support for more model types.}
Previous systems, such as DINC~\cite{zheng2023dinc}, support more models, such as XgBoost~\cite{chen2016xgboost} and isolation forest~\cite{liu2008isolation}. However, they are not runtime-programmable, thus have more flexibility. Supporting these models requires adding new runtime programmable tables to our data plane program. We leave this for future work.

\section{Conclusion}
In this paper, we introduce \sysname{}, an end-to-end system that allows application developers to deploy ML models across programmable data planes in a fully automated way. 
Our extensive evaluation shows that \sysname{} is practical compared to existing state-of-the-art systems. We will make our data plane program, translator, optimizer, and scripts publicly available. 


\bibliographystyle{plain}
\bibliography{acorn}

\clearpage

\appendix
\section{Header Design}\label{appen:header}
The input data and intermediate results will travel between network devices. To facilitate communication, we designed an \sysname{} packet header. 

\begin{figure}[h]
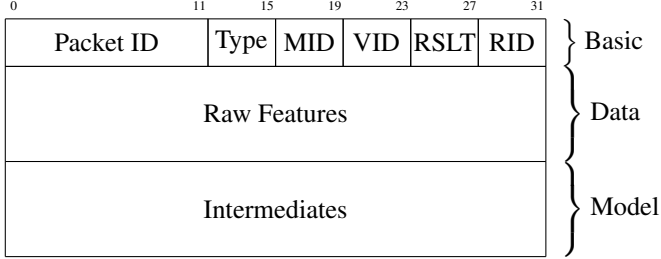

\centering
    \begin{bytefield}[]{32} 
    \bitheader{0, 11, 15, 19, 23, 27, 31} \\ 
    \begin{rightwordgroup}{Basic} 
        \bitbox{12}{Packet ID} & \bitbox{4}{Type} & \bitbox{4}{MID} & \bitbox{4}{VID} & \bitbox{4}{RSLT} & \bitbox{4}{RID}
    \end{rightwordgroup} \\ 
    \begin{rightwordgroup}{Data} 
        \wordbox[tlr]{2}{Raw Features}
    \end{rightwordgroup} \\
    \begin{rightwordgroup}{Model}
        \wordbox[tlrb]{2}{Intermediates}
    \end{rightwordgroup} \\
    \end{bytefield}
\caption{Header design of \sysname{}.}
\label{fig:hdr}
\end{figure}

As figure~\ref{fig:hdr} shows, \sysname{} header consists of two parts. The first part is the basic header. Since multiple packets will run through the network, the basic header has a \textbf{Packet ID} field that distinguishes these packets. This field is vital for the clients as well because the client needs the \textbf{Packet ID} to match the results with raw data. The \textbf{Type} field indicates the type of the packet. There are currently two types of packets, request and response.

There are three fields that store information about the ML model. \textbf{MID}, the model ID, stores the code for the type of ML model. For example, 0 stands for DT, 1 stands for RF, etc. \textbf{VID} is the version information about the model. Since the clients may update their models during runtime, there could be multiple versions of the same model, and we use this field to make sure that the computation is done correctly. \textbf{RSLT} field stores the final prediction result. The \textbf{RID} is the routing code that decides the next destination for this packet.

The second part is the data part, where the raw input data is stored. The size of the data part is decided by the maximum number of supported features, which can be configured by the network operator. The intermediate parts store intermediate results, as we described in \S \ref{sec:trans}.  When the computation finishes, \sysname{}'s data plane program will drop the data part and the intermediates because they are no longer needed. This will reduce the packet size and thus reduce the overhead imposed on the network.

\section{Optimizer: Constraint formulas}
\label{appen:constraints}
\boldpara{Integrity Constraints.} 
\begin{equation*}
\begin{aligned}
       &\sum_{j=1}^{D_s}\sum_{k=1}^{N_D} d_{k}^{p} \cdot x_{ijk} = 1, \; \forall \; i \in [T_s] \\
       &\sum_{i=1}^{N_D} y_i = 1, \; \sum_{j=1}^{D_s} x_{i,j,k} = y_k, \; \forall i \in [T_s], \;\forall k \in [N_D] \\
       &\sum_{i=1}^{N_P}z_{i} = 1
\end{aligned}
\end{equation*}

\boldpara{Guarantee constraints.} 
\begin{equation*}
\begin{aligned}
   d_{k}^{p} \cdot x_{ijk} &\leq y_k, \forall \; i \in [T_s], \forall \; j \in [D_s], \forall \; k \in [N_D] \\
   d_{k}^{p} \cdot x_{ijk} &\leq z_n \; \text{if} \; d_{k} \in P_i, \forall(i,j,k,n) \in [T_s]\times[D_s]\times[N_D]\times[N_P] \\
\end{aligned}
\end{equation*}

\boldpara{Resource limitation.} 
We introduce an indicator, $F_{i, j, k}$, to indicate whether stage $i$ on the program can fit into stage $j$ on $d_k$.
\begin{equation*}
\begin{aligned}
    &\sum_{i=1}^{T_s}\sum_{j=1}^{D_s} d_{k}^p \cdot x_{i,j,k} \leq D_s, \forall k \in [N_D] \\
    &x_{ijk} <= F_{i,j,k}, \;\forall i \in [T_s], \;\forall j \in [N_T^D], \; \forall k \in [N_D]
\end{aligned}
\end{equation*}

\boldpara{Stage dependency.} 
Suppose stage $s_{m}$ is deployed on stage $q$ of device $p_i^{n}$ on an arbitrary path $P_i$. Then, the dependency constraint requires that all predecessor stages of $s_{m}$ must be deployed before $s_m$, on devices $\{p_i^0, p_i^1, \cdots, p_i^n\}$, i.e.,
\begin{equation*}
\begin{aligned}
    d_{f(i,n)}^{p} \cdot x_{m,q,f(i,n)} &\leq \sum_{k=1}^{n-1}\sum_{a=1}^{D_s} d_{f(i,k)}^p \cdot x_{j, a, f(i,k)} + \sum_{b=1}^{q-1} d_{f(i,n)}^p \cdot x_{j, b, f(i,n)} \\
    \forall (i,m,n,q,j) & \in [N_P]\times[T_s]\times[|P_i|]\times[D_s]\times[m-1] \\
\end{aligned}
\end{equation*}
where $f(i, n)$ is a mapping function between $i, n$ and the index of device $p^{n}_{i}$ in $D$.

\boldpara{Position of the last stage.} 
(1) If a path $p_i$ is not chosen, then all $c_{ik}$ must be 0; (2) if the last stage $s_{T_s}$ is not on device $d_k$, then all $c_{ik}$ must be 0. (3) if the last stage $s_{T_s}$ is placed on $d_k$ and $p_i$, then $c_{ij}=1$.
\begin{equation*}
\begin{aligned}
   c_{ik} &\leq z_i, \; \forall i \in [N_P], \;\forall k \in [N_D] \\
   c_{ik} &\leq \sum_{j=1}^{D_s} x_{T_s, j, k}, \; \forall i \in [N_P], \;\forall k \in [N_D] \\
    \sum_{j=1}^{D_s} x_{T_s, j, k} + z_i - c_{ik} &\leq 1, \; \forall i \in [N_P], \forall k \in [N_D] \\
\end{aligned}
\end{equation*}

For random forests and SVMs with multiple hyperplanes, we run the optimizer multiple times, each time for one tree or one hyperplane. We perform trimming to reduce the search space for the optimizer to reduce the solving time.

\section{Additional Information on Evaluations}

\subsection{Datasets}\label{appen:datasets}
We used multiple datasets to evaluate our work. These datasets have been widely used by other works, or represent different types of workloads. In this section, we introduce each dataset briefly and give information about how we preprocess these datasets.

\smallskip
\noindent
\textbf{NSL-KDD.} The NSL-KDD~\cite{nsl-kdd} dataset is a network intrusion detection dataset that is built to improve the classic KDD'99~\cite{kdd99} dataset. In the preprocessing step, we dropped the \texttt{num\_outbound\_cmds} feature as it is the same value for all samples. We also dropped data samples with service: \texttt{aol, harvest, http\_2784, http\_8001, red\_i, urh\_i}, because there are very few data samples. The final dataset has 125,948 training samples and 22,544 testing samples.

\smallskip
\noindent
\textbf{UNSW-IoT.} UNSW-IoT~\cite{unsw-iot} is an IoT device identification dataset that includes traffic flows from a large set of devices. We use 15 days of data for model training and one day for testing, as stated in Jewel~\cite{akem2024jewel}. Our final dataset has 626,463 training samples and 143,141 testing samples.

\smallskip
\noindent
\textbf{CICIDS-17.} The CICIDS-17~\cite{cicids2017} dataset is an intrusion detection dataset designed by the Canadian Institute for Cybersecurity (CIC). This dataset contains over 2.8 million data samples collected over 5 days, and includes normal traffic and 7 types of attack traffic. We used the ``Friday-WorkingHours-Afternoon-DDos'' subset as our input data, dropped all samples with NaN or negative values, and normalized the dataset. We also performed a log1p transformation to make the dataset less skewed. There are 102,996 training samples and 34,333 testing samples after preprocessing.

\smallskip
\noindent
\textbf{UNSW-NB15.} UNSW-NB15~\cite{unswnb15} is a dataset of real normal and contemporary attack behaviours created by the the Cyber Range Lab of UNSW Canberra. We dropped attacks in types: Analysis, Backdoor, Shellcode, Worms, because they have fewer samples compared to other attack types. We also dropped columns that are not available on switches: \texttt{id, service, trans\_depth, response\_body\_len, ct\_flw\_http\_mthd, is\_ftp\_login, ct\_ftp\_cmd, ct\_src\_ltm, ct\_srv\_dst, ct\_srv\_src, ct\_dst\_ltm, ct\_src\_dport\_ltm, ct\_dst\_sport\_ltm, ct\_dst\_src\_ltm, is\_sm\_ips\_ports, stcpb, dtcpb, synack, ackdat}. We also applied a log1p transformation and normalization on the dataset. The final dataset has 175,341 training samples and 75,641 testing samples, and 166 features in total. The targets are normal and attack.


\smallskip
\noindent
\textbf{VCAML} The VCAML~\cite{sharma2023estimating} dataset is a video conferencing dataset collected using three applications: Meet, Webex, and Teams. The dataset has an in-lab subset and a real-world subset. We used the Webex data from the in-lab subset, and extracted 14 features based on the original paper: mean, standard deviation, median, minimum, and maximum of packet sizes and inter-arrival times, and the number of bytes and packets per second for each flow, as well as the number of unique packet sizes observed and the number of microbursts of packets in a window. We use the frame height as the prediction label. The final dataset has 10,011 training samples and 3,371 testing samples.

\smallskip
\noindent
\textbf{Iris.} The Iris~\cite{iris_53} dataset is a classic dataset for building machine learning models. This dataset contains 150 samples of flowers from 3 species, and includes 4 features that describe these flowers. We used the entire dataset in our experiments, with 80\% as training samples, and 20\% as testing samples.

\smallskip
\noindent
\textbf{Digits.} The Digits (Optical Recognition of Handwritten Digits)~\cite{sklearn_digits} dataset contains 5,620 8x8 images of digits from 0 to 9. Each image has 64 pixels, and each pixel is an integer from 0 to 16, representing the normalized greyscale levels. We used 80\% of the dataset as training samples, and 20\% as testing samples.

\smallskip
\noindent
\textbf{MNIST.} The Modified National Institute of Standards and Technology (MNIST)~\cite{mnist} dataset is a database of handwritten digits containing 60,000 training samples and 10,000 testing samples. Each sample is a 28x28 image, making it 784 features in total. This dataset is widely used in machine learning and can be considered as the ``Hello World'' dataset for the ML field. For our experiments, we used 20,000 training samples and 10,000 testing samples.

\smallskip
\noindent
\textbf{SATDAP.} The SATDAP (Predict Students' Dropout and Academic Success)~\cite{student_performance} dataset was created from a higher education institution and was supported by program SATDAP. Each instance in the dataset is a student described by 36 features, and the groundtruth is one of [Enrolled, Dropout, Graduate], making it a three-category dataset. We used 80\% of the dataset as training samples, and 20\% as testing samples.

We summarize the key information about our datasets and present it in Table~\ref{tab:dataset}.

\begin{table}[!h]
\centering
\resizebox{\columnwidth}{!}{
  \begin{tabular}{ccccc}
    \toprule
    \textbf{Dataset} & \textbf{Train Samples} & \textbf{Test Samples} & \textbf{Features} & \textbf{Classes}  \\
    \midrule
    NSL-KDD & 125,948 & 22,544 & 119 & 2 \\
    \midrule
    UNSW-IoT & 626463 & 143141 & 30 & 25  \\
    \midrule
    CICIDS-17 & 102,996 & 34,333 & 78 & 2 \\
    \midrule
    UNSW-NB15 & 175,341 & 75,641 & 166 & 2 \\
    \midrule
    ISCXVPN16 & 2,357 & 590 & 23 & 2 \\
    \midrule
    VCAML & 10011 &3371 & 14 & 2 \\
    \midrule
    Iris & 120 & 30 & 4 & 3 \\
    \midrule
    Digits & 1,437 & 360 & 64 & 10 \\
    \midrule
    MNIST & 20,000 & 10,000 & 784 & 10 \\
    \midrule
    SATDAP & 3,539 & 885 & 36 & 3 \\
    \bottomrule
 \end{tabular}
}
\caption{Information about the dataset setups.}
\label{tab:dataset}
\end{table}

\subsection{Experiment Setups}
\label{appen:exp_setups}
\smallskip
\noindent
\textbf{Data Preprocessing.} Raw input data does not have a fixed range, and it contains both positive and negative numbers, making it difficult to write compatible data plane programs. For every dataset, we used min-max scalers to scale the data into the range of $[0, 1)$ so that it can be processed by the data plane.

\smallskip
\noindent
\textbf{SVM Simulator.} Our simulator mimics the computation flow of SVMs in \sysname{}. For each feature, our simulator looks up the multiplication product in a dictionary and converts the product into signed integers using two's complement representations. The sums are computed on the signed integers, and the simulator extracts the signed bit for each hyperplane as the decision of this hyperplane. The simulator does this for all hyperplanes and makes a prediction using majority voting.

\subsection{SRAM Usage of SVM Models}\label{appen:svmmemory}
\begin{figure}[t]
  \centering
  \includegraphics[width=\linewidth]{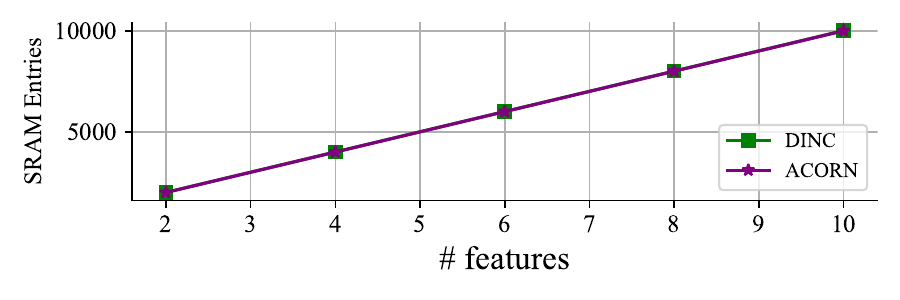}
  \caption{SRAM usage of benchmark systems for SVM.}
  \label{fig:dyn_res_svm}
\end{figure}

With respect to SVM models, our design does not change the way in which SVM models are represented. As Figure~\ref{fig:dyn_res_svm} shows, DINC and \sysname{} use almost the same amount of SRAM for each number of features.


\end{document}